\documentclass[preprint,pra]{revtex4-1}
\usepackage{amsmath}
\usepackage{amssymb}
\usepackage{bm}
\usepackage{xcolor}
\usepackage{graphics}
\usepackage{graphicx}

\begin{document}
\title{Nonrelativistic quantum electrodynamic approach to polarizabilities of light atoms}
\author {Xue-Song Mei$^{1,3}$}
\author {Wan-Ping Zhou$^{2}$}
\author {Hao-Xue Qiao$^{1}$} \thanks{Email address: qhx@whu.edu.cn}

\affiliation {$^1$School of Physics and Technology, Wuhan University, Wuhan 430072, China}
\affiliation {$^2$Engineering and Technology College, Hubei University of Technology, Wuhan 430000, China}
\affiliation {$^3$State Key Laboratory of Magnetic Resonance and Atomic and Molecular Physics, Wuhan Institute of Physics and Mathematics, Chinese Academy of Sciences, Wuhan 430071, China}

\begin{abstract}
We develop a field-quantization scheme for calculating quantum electrodynamic effects on polarizabilities of light atomic systems. This scheme is based on the theory of long-wavelength quantum electrodynamics of Pachucki [Phys. Rev. A \textbf{69}, 052502 (2004)], which combines the theory of nonrelativistic quantum electrodynamics with the Power-Zienau transformation. The external electromagnetic field effects, including electric and magnetic multipole polarizabilities and their relativistic and radiative corrections, are derived using this scheme. The Coulomb-transverse-photon contributions are shown to be zero due to parity symmetry.

\end{abstract}

\maketitle

\section{Intruduction} \label{introduction}
So far, the unified field-quantization method to derive polarizabilities and their various relativistic effects for light atomic systems has been rarely studied. Recently, precision spectroscopies for few-body atomic and molecular systems have provided excellent opportunities for testing quantum electrodynamic (QED) theory and for determining fundamental physical constants, such as the Rydberg constant~\cite{PhysRevLett.78.440, PhysRevLett.110.230801}, the nuclear charge radius of the atom concerned~\cite{PhysRevLett.119.263002},
and the proton-to-electron mass ratio~\cite{2008LNP...745.....K}. The significant discrepancy in the proton charge radius determined from the hydrogen spectroscopy and from the muonic hydrogen spectroscopy has stimulated a lot of research activities, both theoretically and experimentally, on simple atomic systems, demanding for more accurate experimental measurements and more precise theoretical calculations~\cite{Vassen39}.
Although some of the transition frequencies in helium have been measured to an extremely high accuracy, there are still deviations between different experimental results. The helium nuclear charge radius derived from the $2^3S-2^3P$ transition has a $4\sigma$ dispute with the value derived from the $2^3S-2^1S$ transition~\cite{vanrooij2011frequency}. Despite the recent significant improvement over the $2^3S-2^1S$ transition frequency, the discrepancy between the obtained radii is strengthened~\cite{rengelink2018precision}. The $2^3S-2^3P$ transition frequency of $^4$He measured by different experiments still has notable discrepancy~\cite{PhysRevLett.108.143001,PhysRevLett.119.263002}. Experimentally, the blackbody radiation shift due to the AC Stark effect is one of the main systematic errors in some measurements, such as in the development of atomic optical clocks at the level of $10^{-18}$ uncertainty~\cite{PhysRevA.74.020502, PhysRevA.87.012509}. Nowadays, the concepts of magic and tune-out wavelengths are used in order to overcome
the AC Start effects in precision measurements~\cite{2015RvMP...87..637L, PhysRevA.75.053612}. Many calculations have been performed on magic and tune-out wavelengths to provide helpful guidance in searching best candidates for clock systems~\cite{PhysRevA.87.042517, PhysRevA.97.041403}.

Bound-state QED theory for few-electron atomic systems, in particular for helium, has been greatly developed in the past two decades~\cite{PhysRevA.95.062510}, which has become an important cornerstone for precision studies of light atomic systems. For example, the next-highest-order $m\alpha^7$ QED corrections in helium allow for a direct extraction of the nuclear radius of $^4$He~\cite{PhysRevA.98.032503}. Calculations on polarizabilities of helium isotopes have focussed on the magic wavelength and tune-out wavelength for the transitions $2^3S-2^3P$ and $2^1S-2^3S$ and the involved states~\cite{PhysRevA.88.052515,PhysRevA.90.052508, PhysRevA.93.052516} and a good agreement with experiment has been achieved~\cite{PhysRevA.98.040501,rengelink2018precision}. Relativistic effects on dynamical polarizabilities have become non-negligible, and radiative corrections will become more and more important as experimental precision is improving. There are works on relativistic and QED corrections to the static polarizability~\cite{PhysRevA.63.012504, PhysRevLett.92.233001} and relativistic corrections to the dynamical
polarizability~\cite{PhysRevLett.114.173004} of helium. In these calculations, relativistic corrections were derived by perturbing the known nonrelativistic polarizability, and the radiative corrections were derived by taking second-order electric-field derivative of the Bethe-logarithm term. In contrast to the growing importance of relativistic and radiative effects in atoms in an external electromagnetic field, a unified field-quantization scheme for treating dynamical polarizability is still less investigated, including contributions from magnetic dipole, electric quadrupole, and higher order effects.

The purpose of this paper is to present a unified scheme, based on the long-wavelength QED theory of Pachucki~\cite{PhysRevA.69.052502}, for studying the polarizabilities of a light atom and their relativistic and radiative corrections, where the external electromagnetic fields are kept at second order. Combining the theory of nonrelativistic quantum electrodynamics (NRQED) with the Power-Zienau (PZ) transformation, the dynamical polarizability and its relativistic and radiative corrections are obtained under this unified framework. We introduce the principle of this scheme in Sec.~\ref{theory and method}. Dynamical polarizability and its relativistic and radiative corrections are treated in Sec.~\ref{dynamical polarizability}. In Sec.~\ref{magnetic polarizability}, we introduce the concept of magnetic polarizability and its relativistic corrections.
In Sec.~\ref{other types of corrections}, we discuss other types of corrections, including the one from the Coulomb-transverse-photon.
Two notes on the PZ transformation and radiative calculations are contained in Appendices ~\ref{appendix_note_1} and ~\ref{appendix_note_2}.
Throughout this paper, the Einstein summation convention is adopted, where we do not distinguish the covariant and contravariant variables, {\it i.e.}, $x^i=x_i$. Also the system of natural units is used where $\hbar=1$ and $c=1$.

\section{Theory and method} \label{theory and method}

NRQED is an effective field theory on light atomic systems that is based on an effective Lagrangian of nonrelativistic field. This theory was first introduced by Caswell and Lepage~\cite{caswell1986effective}. Another form of NRQED was developed by Pachucki~\cite{PhysRevA.71.012503,PhysRevA.82.052520}, which is based on an effective Hamiltonian derived from the Foldy-Wouthuysen transformation. The singularity of Green's function determines the energy-shift.

For an electron in an electromagnetic field, the relativistic effects can be described by the well-known Dirac Hamiltonian
\begin{equation}
\begin{aligned}
H_D = \bm{\alpha}\cdot \bm{\pi}+\beta m + eA^0,
\end{aligned}
\end{equation}
where $\bm{\pi}=\bm{p}-e\bm{A}$, with $A^\mu$ being the electromagnetic four potential, and $\bm{\alpha}$ and $\bm{\pi}$ are usual Dirac matrices.
The energy shift due to the external electromagnetic field to second order can be written in the form
\begin{equation}\label{Energy shift}
\begin{aligned}
\Delta E=
\sum_{a,b}
\left\langle \psi_D \left|
\mathcal{J}_a^{i} e^{\imath \bm{k}\cdot\bm{x}_a}
\frac{1}{E-H_D + \omega}
\mathcal{J}_b^{j\dag} e^{{-\imath} \bm{k}\cdot\bm{x}_b}
+
\mathcal{J}_a^{i\dag} e^{\imath \bm{k}\cdot\bm{x}_a}
\frac{1}{E-H_D - \omega}
\mathcal{J}_b^{j} e^{{-\imath} \bm{k}\cdot\bm{x}_b}
\right| \psi_D \right\rangle,
\end{aligned}
\end{equation}
where $\mathcal{J}$ is the electromagnetic current operator that describes electromagnetic interaction, $|\psi_D\rangle$ is an eigenstate of the Dirac Hamiltonian $H_D$, and the summation over $a$ and $b$ represents different fermions in the system. Generally, for a light atomic system the
energy levels can be determined approximately by solving the corresponding Schr\"odinger equation because of the characteristic of low momenta of the electrons. The relativistic and radiative corrections can then be taken into account using perturbation theory. In order to make Eq.~\eqref{Energy shift} suitable for accommodating Schr\"odinger wave functions, one should perform a nonrelativistic approximation to reveal relativistic corrections. The common step towards this is to perform the Foldy-Wouthuysen (FW) transformation~\cite{PhysRev.78.29},
which is a unitary transformation defined by
\begin{equation}
\begin{aligned}
H_{\rm FW}=\text{e}^{{\imath} S} (H_D-{\imath} \partial_t) \text{e}^{-{\imath} S},
\end{aligned}
\end{equation}
where $S$ is the transformation operator that is not unique. One can choose the simplest form
\begin{equation}
\begin{aligned}
S=\frac{- \imath}{2m} \beta (\bm{\alpha}\cdot \bm{\pi}),
\end{aligned}
\end{equation}
or choose a more complicated form~\cite{PhysRevA.71.012503,0256-307X-31-6-063102}
\begin{equation}
\begin{aligned}
S&=\frac{-\imath}{2m}\left\{ \beta \bm{\alpha} \cdot \bm{\pi}+\frac{-\beta}{3m^2}\left( \bm{\alpha} \cdot \bm{\pi} \right) ^3+\frac{1}{2m}\left[ \bm{\alpha} \cdot \bm{\pi},eA^0-\imath\partial _t \right] \right\},
\end{aligned}
\end{equation}
which could bring some convenience in higher-order calculations. However, no matter how we choose the operator $S$, the resulting FW Hamiltonian should be equivalent at a given order of magnitude. Here we give the FW Hamiltonian accurate up to the order of $m\alpha^6$, with $\alpha$ being the fine-structure constant
\begin{equation}\label{FWHamil}
\begin{aligned}
\mathcal{H}_{\rm FW}
&=
\frac{(\bm{\sigma}\cdot\bm{\pi})^2}{2m}
+eA^0
-\frac{(\bm{\sigma}\cdot\bm{\pi})^4}{8m^3}
-\frac{{\imath}}{8m^2} [\bm{\sigma}\cdot\bm{\pi}, \bm{\sigma}\cdot\bm{\mathcal{E}}]
\\&
+\frac{(\bm{\sigma}\cdot\bm{\pi})^6}{16m^5}
+\frac{3{\imath}}{64m^4}
\{(\bm{\sigma}\cdot\bm{\pi})^2,[\bm{\sigma}\cdot\bm{\pi}, \bm{\sigma}\cdot\bm{\mathcal{E}}]\}
\\&
+\frac{5{\imath}}{128m^4}
[(\bm{\sigma}\cdot\bm{\pi})^2, \{\bm{\sigma}\cdot\bm{\pi}, \bm{\sigma}\cdot\bm{\mathcal{E}}\}]
+\frac{1}{8m^3} \bm{\mathcal{E}}^2
+\mathcal{O}(m\alpha^8)\,,
\end{aligned}
\end{equation}
where the use of calligraphic style $\mathcal{H}_{\rm FW}$ is to emphasize that it is a Hamiltonian density.

The FW transformation allows one to obtain the nonrelativistic form of the current operator $\mathcal{J}$. Also, the nonrelativistic expansions for the relativistic wave function and Green's function are respectively
\begin{equation}
\begin{aligned}
|\psi_D\rangle \rightarrow \left(1 + \frac{Q}{E_0-H_0}H' + \cdots\right)
|\phi_0\rangle
,
\notag
\end{aligned}
\end{equation}
\begin{equation}
\begin{aligned}
\frac{1}{E-H_D\pm\omega}
&\rightarrow \frac{1}{E_0 - H_0\pm\omega }
- \frac{1}{E_0-H_0\pm\omega}(E'-H')\frac{1}{E_0-H_0\pm\omega}
+ \cdots,
\end{aligned}
\end{equation}
where $H_0$ is the Schr\"odinger-Coulomb Hamiltonian for the system concerned, $|\phi_0\rangle$ is the eigenstate of $H_0$, and $Q/(E_0-H_0)$ is
the reduced Green's function with $Q=1-|\phi_0\rangle \langle \phi_0 |$ the projection operator. Also in the above, $H'$ is the perturbing Hamiltonian that has a different form for different situation, such as the one with relativistic or radiative origin, and $E'$ is the expectation value of $H'$. Finally, $\omega$ is the energy of photon from the external electromagnetic field.
It is noted that $e^{\imath\bm{k}\cdot\bm{x}}\sim 1$ under the long-wavelength condition $|\bm{k}\cdot \bm{x}|\ll 1$.

With the FW transformation and the nonrelativistic expansions introduced above, one can properly describe relativistic and quantum-field effects on a nonrelativistic atomic system. It should be noted that the electromagnetic four potential $A^\mu$ in the Dirac Hamiltonian should contain both the interaction for the atom itself and the interaction from the external electromagnetic field,
which may lead to some difficulties for radiative calculations. In general, the external electromagnetic field satisfies the long-wavelength condition, meaning that the wavelength of the external photon is much larger than the Bohr radius. We rewrite the four potential as $A^\mu=A^\mu_{\text{at}}+A^\mu_{\text{ex}}$, where $A^\mu_{\text{at}}$ is the field from within the atom and $A^\mu_{\text{ex}}$ is due to the external field. With this distinction, we can replace $\bm{\pi}$ by $\bm{\Pi}=\bm{p}-e(\bm{A}_{\text{at}}+\bm{A}_{\text{ex}})$ in the FW Hamiltonian so that it contains these two fields.

In order to distinguish different effects from the two fields and extract contributions from multipole moments, we carry out another unitary
transformation called the Power-Zienau transformation~\cite{1959RSPTA.251..427P}.  In general, the Power-Zienau transformation on an Hamiltonian $H$ is defined by
\begin{equation}
\begin{aligned}
H_{\rm PZ}=e^{-\imath\phi}He^{\imath\phi}+\partial _t\phi\,.
\end{aligned}
\end{equation}
The phase $\phi$ here is
\begin{equation}\label{phase def}
\begin{aligned}
\phi =e\int_0^1{\bm{r}\cdot \bm{A}_{\text{ex}}\left( u\bm{r} \right) \text{d}u}.
\end{aligned}
\end{equation}
Under the long-wavelength condition, the electromagnetic four potential of the external field can be expanded into a multipole form
\begin{equation}\label{A multipole form}
\begin{aligned}
A_{\text{ex}}^{\mu}\left( \bm{r},t \right) =A_{\text{ex}}^{\mu}\left( \text{0,}t \right) +r^iA_{\text{ex}}^{\mu}\left( \text{0,}t \right) _{,i}+\frac{1}{\text{2!}}r^ir^jA_{\text{ex}}^{\mu}\left( \text{0,}t \right) _{,ij}+\frac{1}{\text{3!}}r^ir^jr^kA_{\text{ex}}^{\mu}\left( \text{0,}t \right) _{,ijk}+\cdots,
\\
\end{aligned}
\end{equation}
where $
A_{\text{ex}}^{\mu}\left( \text{0,}t \right) _{,i}\equiv \frac{\partial A_{\text{ex}}^{\mu}\left( \text{0,}t \right)}{\partial r^i}.
$
Substituting \eqref{A multipole form} into \eqref{phase def}, we have
\begin{equation}\label{A_multipole}
\begin{aligned}
\phi =e\left[ r^iA_{\text{ex}}^{i}\left( \text{0,}t \right) +\frac{1}{\text{2!}}r^ir^jA_{\text{ex}}^{i}\left( \text{0,}t \right) _{,j}+\frac{1}{\text{3!}}r^ir^jr^kA_{\text{ex}}^{i}\left( \text{0,}t \right) _{,jk}+\frac{1}{\text{4!}}r^ir^jr^kr^lA_{\text{ex}}^{i}\left( \text{0,}t \right) _{,jkl}+\cdots \right]
\end{aligned}
\end{equation}
and
\begin{equation}\label{multipole_indi}
\begin{aligned}
eA^0_{\text{ex}}+\partial _t\phi =-e\left[ \bm{r}\cdot \bm{\mathcal{E}}_{\text{ex}}\left( \text{0,}t \right) +\frac{1}{\text{2!}}r^ir^j\mathcal{E}_{\text{ex}}^{i}\left( \text{0,}t \right) _{,j}+\frac{1}{\text{3!}}r^ir^jr^k\mathcal{E}_{\text{ex}}^{i}\left( \text{0,}t \right) _{,jk}+\cdots \right]\,.
\end{aligned}
\end{equation}

Considering the leading-order relativistic corrections, we use the FW Hamiltonian up to order $m\alpha^4$
\begin{equation}\label{reversed_FW_hami}
\begin{aligned}
\mathcal{H}_{\rm FW}=\frac{\left( \boldsymbol{\sigma }\cdot \mathbf{\Pi } \right) ^2}{2m}+e\left( A_{\text{at}}^{0}+A_{\text{ex}}^{0} \right) -\frac{\left( \boldsymbol{\sigma }\cdot \mathbf{\Pi } \right) ^4}{8m^3}-\frac{\imath}{8m^2}\left[ \boldsymbol{\sigma }\cdot \mathbf{\Pi ,}\boldsymbol{\sigma }\cdot \bm{\mathcal{E}}_{\text{at}} \right]\,.
\end{aligned}
\end{equation}
We then perform the following PZ transformation
$$
\mathcal{H}_{\rm PZ}=e^{-\imath\phi}\mathcal{H}_{\rm FW}e^{\imath\phi}+\partial _t\phi\,,
$$
which gives rise to
\begin{equation}
\begin{aligned}
\mathcal{H}_{\rm PZ}&=\frac{1}{2m}\left[ \bm{\sigma }\cdot \left( \bm{\pi }+\frac{e}{2}\bm{r}\times \bm{\mathcal{B}}_{\text{ex}} \right) \right] ^2-e\left[ \bm{r}\cdot \bm{\mathcal{E}}_{\text{ex}}\left( \text{0,}t \right) +\frac{1}{\text{2!}}r^ir^j\mathcal{E}_{\text{ex}}^{i}\left( \text{0,}t \right) _{,j}+\frac{1}{\text{3!}}r^ir^jr^k \mathcal{E}_{\text{ex}}^{i}\left( \text{0,}t \right) _{,jk}+\cdots \right]
\\&
+\frac{-1}{8m^3}\left[ \bm{\sigma }\cdot \left( \bm{\pi }+\frac{e}{2}\bm{r}\times \bm{\mathcal{B}}_{\text{ex}} \right) \right] ^4+\frac{-\imath}{8m^2}\left[  \bm{\sigma }\cdot \left( \bm{\pi }+\frac{e}{2}\bm{r}\times \bm{\mathcal{B}}_{\text{ex}} \right) ,\bm{\sigma }\cdot \bm{\mathcal{E}}_{\text{at}} \right] ,
\end{aligned}
\end{equation}
where $\bm{\mathcal{E}}_{\text{ex}}$ and $\bm{\mathcal B}_{\text{ex}}$ represent, respectively, the external electric and magnetic field, and once again  $\bm{\pi}=\bm{p}-e\bm{A}_{\text{at}}$. Here we have used the following approximation
$$
e^{-\imath\phi}\bm{\Pi}e^{\imath\phi}\simeq \bm{p}+\left[ -\imath\phi ,\bm{p} \right] -e\bm{A}_{\text{at}}-e\bm{A}_{\text{ex}}=\bm{p}-e\bm{A}_{\text{at}}+ \nabla \phi  -e\bm{A}_{\text{ex}}.
$$
Some calculational details can be found in Appendix~\ref{appendix_note_1}.
$\mathcal{H}_{PZ}$ can further be categorized into the following four different types of interactions
$$
\mathcal{H}_{\rm PZ}=
\mathcal{H}_{\text{at}}
+\mathcal{H}_{\rm E}
+\mathcal{H}_{\rm M}
+\mathcal{H}_{\rm EM},
$$
where $\mathcal{H}_{\text{at}}$ represents the interaction from within the atom, $\mathcal{H}_{\rm E}$ is the interaction from the external electric field, $\mathcal{H}_{\rm M}$ is the interaction from the external magnetic field, and $\mathcal{H}_{\rm EM}$ is the term caused by the coupling between the external electric and magnetic fields. These Hamiltonians can be expressed according to
\begin{equation}\label{atomic_hami}
\begin{aligned}
&
\mathcal{H}_{\text{at}}
=eA^0_{\text{at}}+\frac{1}{2m}\left( \bm{\sigma}\cdot \bm{\pi} \right) ^2+\frac{-1}{8m^3}\left( \bm{\sigma}\cdot \bm{\pi} \right) ^4+\frac{-\imath}{8m^2}\left[ \bm{\sigma}\cdot \bm{\pi},\bm{\sigma}\cdot \bm{\mathcal{E}}_{\text{at}} \right],
\end{aligned}
\end{equation}

\begin{equation}\label{electric_hami}
\begin{aligned}
\mathcal{H}_{\rm E}=-e\left[ \bm{r}\cdot \bm{\mathcal{E}}_{\text{ex}}\left( \text{0,}t \right) +\frac{1}{\text{2!}}r^ir^j\mathcal{E}_{\text{ex}}^{i}\left( \text{0,}t \right) _{,j}+\frac{1}{\text{3!}}r^ir^jr^k\mathcal{E}_{\text{ex}}^{i}\left( \text{0,}t \right) _{,jk}+\cdots \right] +\frac{-\imath}{8m^2}\left[ \bm{\sigma}\cdot \bm{\pi},\bm{\sigma}\cdot \bm{\mathcal{E}}_{\text{ex}} \right],
\end{aligned}
\end{equation}

\begin{equation}\label{H_magnetic_hami}
\begin{aligned}
\mathcal{H}_{\rm M}&=
\frac{1}{2m}\left\{  \boldsymbol{\sigma }\cdot \boldsymbol{\pi },\boldsymbol{\sigma }\cdot \frac{e}{2}\boldsymbol{r}\times \bm{\mathcal{B}}_{\text{ex}} \right\}
+\frac{-\imath}{8m^2}\left[ \boldsymbol{\sigma }\cdot \frac{e}{2}\boldsymbol{r}\times \bm{\mathcal{B}}_{\text{ex}},\boldsymbol{\sigma }\cdot \mathcal{E}_{\text{at}} \right]
\\&
+\frac{-1}{8m^3}\left\{ \left( \boldsymbol{\sigma }\cdot \boldsymbol{\pi } \right) ^3,\boldsymbol{\sigma }\cdot \frac{e}{2}\boldsymbol{r}\times \bm{\mathcal{B}}_{\text{ex}}  \right\} +\frac{-1}{8m^3}\left( \boldsymbol{\sigma }\cdot \boldsymbol{\pi } \right) \left\{  \boldsymbol{\sigma }\cdot \boldsymbol{\pi }  , \boldsymbol{\sigma }\cdot \frac{e}{2}\boldsymbol{r}\times \bm{\mathcal{B}}_{\text{ex}}  \right\} \left( \boldsymbol{\sigma }\cdot \boldsymbol{\pi } \right)
\\&
+\frac{1}{2m}\left( \boldsymbol{\sigma }\cdot \frac{e}{2}\boldsymbol{r}\times \bm{\mathcal{B}}_{\text{ex}} \right) ^2+\frac{-1}{8m^3}\left\{ \left( \boldsymbol{\sigma }\cdot \boldsymbol{\pi } \right) ^2,\left( \boldsymbol{\sigma }\cdot \frac{e}{2}\boldsymbol{r}\times \bm{\mathcal{B}}_{\text{ex}} \right) ^2 \right\}
\\&
+\frac{-1}{8m^3}\left\{  \boldsymbol{\sigma }\cdot \boldsymbol{\pi }  , \boldsymbol{\sigma }\cdot \frac{e}{2}\boldsymbol{r}\times \bm{\mathcal{B}}_{\text{ex}}  \right\} ^2+\frac{-1}{8m^3}\left\{ \boldsymbol{\sigma }\cdot \boldsymbol{\pi }  ,\left( \boldsymbol{\sigma }\cdot \frac{e}{2}\boldsymbol{r}\times \bm{\mathcal{B}}_{\text{ex}} \right) ^3 \right\}
\\&
+\frac{-1}{8m^3}\left( \boldsymbol{\sigma }\cdot \frac{e}{2}\boldsymbol{r}\times \bm{\mathcal{B}}_{\text{ex}} \right) \left\{  \boldsymbol{\sigma }\cdot \boldsymbol{\pi } , \boldsymbol{\sigma }\cdot \frac{e}{2}\boldsymbol{r}\times \bm{\mathcal{B}}_{\text{ex}}  \right\} \left( \boldsymbol{\sigma }\cdot \frac{e}{2}\boldsymbol{r}\times \bm{\mathcal{B}}_{\text{ex}} \right)
\\&
+\frac{-1}{8m^3}\left( \boldsymbol{\sigma }\cdot \frac{e}{2}\boldsymbol{r}\times \bm{\mathcal{B}}_{\text{ex}} \right) ^4,
\end{aligned}
\end{equation}
\begin{equation}\label{H_electric_magnetic_hami}
\begin{aligned}
\mathcal{H}_{\rm EM}=\frac{-\imath}{8m^2}\left[ \bm{\sigma}\cdot \frac{e}{2}\bm{r}\times \bm{\mathcal{B}}_{\text{ex}},\bm{\sigma}\cdot \bm{\mathcal{E}}_{\text{ex}} \right].
\end{aligned}
\end{equation}
With these Hamiltonians, the interaction vertices can be created and the external electromagnetic field effects can thus be evaluated more explicitly. For example, the electric polarizability of atom is the second-order effect of the external electric field, which can be attributed to the Hamiltonian $\mathcal{H}_{\rm E}$. Similarly, the magnetic polarizability can be attributed to the Hamiltonian $\mathcal{H}_{\rm M}$.
It should be mentioned that,
when we talk about the radiative correction of an atom, it is due to the virtual photons of the atomic Hamiltonian $\mathcal{H}_{\text{at}}$. Remember that our calculations are performed under the long-wavelength condition, which implies that the length gauge is used for the external field interaction. For the virtual photons, we use the velocity gauge in the corresponding Hamiltonian.

\section{Dynamical Polarizability} \label{dynamical polarizability}
The dynamical polarizability of an atom is a measure of response to an AC Stark effect. The shift of an energy level of atom in a weak
electric field $\mathcal{E}$ can be expressed
in the form using perturbation theory
\begin{equation}\label{def_polarizability}
\begin{aligned}
\Delta E = \frac{-1}{2} {\mathcal{E}}^{i*}_{\text{ex}} \mathcal{E}^j_{\text{ex}} \alpha^{ij},
\end{aligned}
\end{equation}
where $\alpha^{ij}$ is the dipole polarizability tensor by definition, which can further be decomposed into symmetric, anti-symmetric, and symmetric-no-trace parts
\begin{equation}\label{sym_form}
\begin{aligned}
\mathcal{E}^{i\dag}_{\text{ex}} \mathcal{E}^j_{\text{ex}} \alpha ^{ij}
&=
\mathcal{E}_{\text{ex}}^{k}\mathcal{E}_{\text{ex}}^{k\dag}\alpha _{\rm S}-{\imath}\mathcal{E}_{\text{ex}}^{i}\mathcal{E}_{\text{ex}}^{j\dag}\alpha _{\rm V}^{k}+\mathcal{E}_{\text{ex}}^{i}\mathcal{E}_{\text{ex}}^{j\dag}\alpha _{\rm T}^{ij},
\end{aligned}
\end{equation}
where
\begin{equation}\label{sym_form2}
\begin{aligned}
\alpha _{\rm S}&=\frac{\delta ^{ij}}{3}\alpha ^{ij}
,~
\alpha _{\rm V}^{k}
&=\imath\frac{\epsilon ^{ijk}}{2}\alpha ^{ij}
,~
\alpha _{\rm T}^{ij}
&=\frac{\alpha ^{ij}+\alpha ^{ji}}{2}-\frac{\delta ^{ij}}{3}\alpha ^{ij}
\end{aligned}
\end{equation}
are respectively called the scalar, the vector, and the tensor parts of the polarizability. In the
above $\epsilon^{ijk}$ is the Levi-Civita symbol.

\subsection{Nonrelativistic Dynamical Polarizability}
The external electric Hamiltonian $\mathcal{H}_{\rm E}$ is given in Eq.~(\ref{electric_hami}), 
in which the first term in the first square brackets stands for the electric dipole interaction that causes the following energy shift
\begin{equation}
\begin{aligned}
\delta H_{\rm E,NR}=\sum_{a,b}{}\left( -er_{a}^{i}\mathcal{E}_{\text{ex}}^{i} \right) G_0\left( -\omega \right) \left( -e\mathcal{E}_{\text{ex}}^{j\dag}r_{b}^{j} \right) +\left( -er_{a}^{i}\mathcal{E}_{\text{ex}}^{i\dag} \right) G_0\left( \omega \right) \left( -e\mathcal{E}_{\text{ex}}^{j}r_{b}^{j} \right)\,.
\end{aligned}
\end{equation}
This expression can further be recast into the form
\begin{eqnarray}\label{NR-energy}
\delta H_{\rm E,NR}=e^2\sum_{a,b}{}\mathcal{E}_{\text{ex}}^{i}\mathcal{E}_{\text{ex}}^{j\dag}\left[ r_{a}^{i} G_0(-\omega) r_{b}^{j}+r_{a}^{j} G_0(\omega) r_{b}^{i} \right]\,,
\end{eqnarray}
where $G_0(\pm\omega)=1/(E_0-H_0\pm\omega)$ is the Green's function.
According to the definition of polarizability Eq.~\eqref{def_polarizability}, we can extract a general expression for the nonrelativistic dynamical polarizability operator
\begin{equation}
\begin{aligned}
\hat{\alpha}^{ij}_{\rm NR}
=
-2e^2
\sum_{a,b}
\left[ r_{a}^{i}G_0(-\omega)r_{b}^{j}+r_{a}^{j}G_0(\omega)r_{b}^{i} \right]\,,
\end{aligned}
\end{equation}
from which we can obtain the following scalar, vector, and tensor parts of the polarizability operator according to Eqs.~\eqref{sym_form} and \eqref{sym_form2}
\begin{eqnarray}
\label{alpha_S}
\hat{\alpha} _{\rm S,NR}&=&
-\frac{4}{3}\delta ^{ij}e^2\sum_{a,b}{}r_{a}^{i}\frac{\left( E_0-H_0 \right)}{\left( E_0-H_0 \right) ^2-\omega ^2}r_{b}^{j},\\
\label{alpha_V}
\hat{\alpha} _{\rm V,NR}^{k}
&=&-2{\imath}\epsilon ^{ijk}e^2\sum_{a,b}{}r_{a}^{i}\frac{\omega}{\left( E_0-H_0 \right) ^2-\omega ^2}r_{b}^{j},\\
\label{alpha_T}
\hat{\alpha} _{\rm T,NR}^{ij}
&=&e^2\sum_{a,b}{}r_{a}^{\left\{ i \right.}\frac{-2\left( E_0-H_0 \right)}{\left( E_0-H_0 \right) ^2-\omega ^2}r_{b}^{j\}}+\frac{4}{3}\delta ^{ij}e^2\sum_{a,b}{}r_{a}^{i}\frac{\left( E_0-H_0 \right)}{\left( E_0-H_0 \right) ^2-\omega ^2}r_{b}^{j}\,,
\end{eqnarray}
where the curly braces stand for $A^{\{i} \hat{O} B^{j\}}=\frac{1}{2}(A^i \hat{O} B^j+A^j \hat{O} B^i)$ for any operator $\hat{O}$ sandwiched between $A$ and $B$.

\subsection{Leading-Order Relativistic Corrections}
The term $\frac{-\imath}{8m^2}\left[ \bm{\sigma}\cdot \bm{\pi},\bm{\sigma}\cdot \bm{\mathcal{E}}_{\text{ex}} \right]$ in $\mathcal{H}_{\rm E}$ of Eq.~(\ref{electric_hami}) represents the leading-order relativistic effect induced by the external electric field. The corresponding leading-order relativistic contribution to the energy is thus given by
\begin{equation}\label{rel_correction}
\begin{aligned}
\delta H_{\rm E,R}&
=H_{\rm R}\frac{Q}{E_0-H_0}\delta H_{\rm E,NR}+\delta H_{\rm E,NR}\frac{Q}{E_0-H_0}H_{\rm R}
\\&
+\sum_{a,b}{}\left[ \left( -er_{a}^{i}\mathcal{E}_{\text{ex}}^{i} \right) G^{\left( 1 \right)}\left( -\omega \right) \left( -er_{b}^{j}\mathcal{E}_{\text{ex}}^{j\dag} \right) +\left( -er_{a}^{i}\mathcal{E}_{\text{ex}}^{i\dag} \right) G^{\left( 1 \right)}\left( \omega \right) \left( -er_{b}^{j}\mathcal{E}_{\text{ex}}^{j} \right) \right]
\\&
+\sum_{a,b}{}\left( -er_{a}^{i}\mathcal{E}_{\text{ex}}^{i} \right) \frac{-\omega}{E_0-H_0-\omega}\frac{1}{8m_b}\left[ \boldsymbol{\sigma }\cdot \boldsymbol{r,\sigma }\cdot \bm{\mathcal{E}}_{\text{ex}}^{\dag} \right] _b
\\&
-\sum_{a,b}{}\frac{1}{8m_a}\left[ \boldsymbol{\sigma }\cdot \bm{r},\boldsymbol{\sigma }\cdot \bm{\mathcal{E}}_{\text{ex}} \right] _a\frac{-\omega}{E_0-H_0-\omega}\left( -er_{b}^{j}\mathcal{E}_{\text{ex}}^{j\dag} \right)
\\&
+\sum_{a,b}{}\left( -er_{a}^{i}\mathcal{E}_{\text{ex}}^{i\dag} \right) \frac{\omega}{E_0-H_0+\omega}\frac{1}{8m_b}\left[ \boldsymbol{\sigma }\cdot \boldsymbol{r,\sigma }\cdot \bm{\mathcal{E}}_{\text{ex}} \right] _b
\\&
-\sum_{a,b}{}\frac{1}{8m_a}\left[ \boldsymbol{\sigma }\cdot \boldsymbol{r,\sigma }\cdot \bm{\mathcal{E}}_{\text{ex}}^{\dag} \right] _a\frac{\omega}{E_0-H_0+\omega}\left( -er_{b}^{j}{\mathcal{E}}_{\text{ex}}^{j} \right) ,
\end{aligned}
\end{equation}
where $H_{\rm R}$ is the Breit-Pauli Hamiltonian~\cite{PhysRevA.74.022512, bethe1957quantum}
\begin{equation}\label{Breit-Pauli Hamiltonian}
\begin{aligned}
H_{\rm R}&=\sum_a{}\left\{ -\frac{\boldsymbol{p}_{a}^{4}}{8m^3}+\frac{\pi Z\alpha}{2m^2}\delta ^3\left( {\boldsymbol r}_a \right) +\frac{Z\alpha}{4m^2}\boldsymbol{\sigma }_a\cdot \frac{\boldsymbol{r}_a}{r_{a}^{3}}\times \boldsymbol{p}_a \right\}
\\&
+\sum_{a,b}{}\left\{ -\frac{\pi \alpha}{m^2}\delta ^3\left( {\boldsymbol r}_{ab} \right) -\frac{\alpha}{2m^2}p_{a}^{i}\left( \frac{\delta ^{ij}}{r_{ab}}+\frac{r_{ab}^{i}r_{ab}^{j}}{r_{ab}^{3}} \right) p_{b}^{j} \right.
\\&
-\frac{2\pi \alpha}{3m^2}\boldsymbol{\sigma }_a\cdot \boldsymbol{\sigma }_b\delta ^3\left( {\boldsymbol r}_{ab} \right) +\frac{\alpha}{4m^2}\frac{\sigma _{a}^{i}\sigma _{b}^{j}}{r_{ab}^{3}}\left( \delta ^{ij}-3\frac{r_{ab}^{i}r_{ab}^{j}}{r_{ab}^{2}} \right)
\\&
\left. +\frac{\alpha}{2m^2r_{ab}^{3}}\left[ \left( \boldsymbol{\sigma }_a\cdot \boldsymbol{r}_{ab}\times \boldsymbol{p}_b-\boldsymbol{\sigma }_b\cdot \boldsymbol{r}_{ab}\times \boldsymbol{p}_a \right) +\left( \boldsymbol{\sigma }_b\cdot \boldsymbol{r}_{ab}\times \boldsymbol{p}_b-\boldsymbol{\sigma }_a\cdot \boldsymbol{r}_{ab}\times \boldsymbol{p}_a \right) \right] \right\} ,
\end{aligned}
\end{equation}
and the first-order expansion of the Green's function is 
\begin{eqnarray}\label{green_G1}
G^{(1)}(\pm \omega)=-G_0(\pm\omega)(E_{\rm R}-H_{\rm R})G_0(\pm\omega)\,,
\end{eqnarray}
with $E_{\rm R}=\langle H_{\rm R}\rangle$.
Also since we only consider the leading-order relativistic correction, we set $\left[ \bm{\sigma}\cdot \bm{\pi},\bm{\sigma}\cdot \bm{\mathcal{E}}_{\text{ex}} \right]\simeq \left[\bm{\sigma}\cdot\bm{p}, \bm{\sigma} \cdot \bm{\mathcal{E}}_{\text{ex}} \right]$.
We can see that the last four terms in Eq.~\eqref{rel_correction}, which contain the commutator $[\bm\sigma \cdot \bm r, \bm\sigma \cdot \bm{\mathcal{E}}_{\rm{ex}}]$, representing the propagation between the electric dipole interaction and the leading-order relativistic interaction, have only anti-symmetric part, {\it i.e.}, the vector part.
Therefore, the symmetrized polarizability operators are 
\begin{equation}\label{rel_scalar}
\begin{aligned}
\hat{\alpha}_{S,R}&=\frac{-2e^2\delta ^{ij}}{3}\sum_{a,b}{}\left[ H_R\frac{Q}{E_0-H_0}r_{a}^{i}\frac{2\left( E_0-H_0 \right)}{\left( E_0-H_0 \right) ^2-\omega ^2}r_{b}^{j}+r_{a}^{i}\frac{2\left( E_0-H_0 \right)}{\left( E_0-H_0 \right) ^2-\omega ^2}r_{b}^{j}\frac{Q}{E_0-H_0}H_R \right]
\\&
+\frac{-2e^2\delta ^{ij}}{3}\sum_{a,b}{}\left[ r_{a}^{i}G^{\left( 1 \right)}\left( -\omega \right) r_{b}^{j}+r_{a}^{i}G^{\left( 1 \right)}\left( \omega \right) r_{b}^{j} \right],
\end{aligned}
\end{equation}
\begin{equation}\label{rel_vector}
\begin{aligned}
\hat{\alpha}_{V,R}^{k}&=
-\text{2}\imath e^2\omega \epsilon ^{ijk}\sum_{a,b}{}\left[ H_R\frac{1}{E_0-H_0}r_{a}^{i}\frac{1}{\left( E_0-H_0 \right) ^2-\omega ^2}r_{b}^{j}+r_{a}^{i}\frac{1}{\left( E_0-H_0 \right) ^2-\omega ^2}r_{b}^{j}\frac{1}{E_0-H_0}H_R \right]
\\&
+\imath e^2\epsilon ^{ijk}\sum_{a,b}{}\left[ r_{a}^{i}G^{\left( 1 \right)}\left( -\omega \right) r_{b}^{j}-r_{a}^{i}G^{\left( 1 \right)}\left( \omega \right) r_{b}^{j} \right]
\\&
+\frac{-e\omega}{4m}\epsilon ^{ijk}\sum_{a,b}{}\left\{ \left[ r_{a}^{i}G_0\left( -\omega \right) \left( \sigma \times {r}_b \right) ^j+r_{a}^{i}G_0\left( \omega \right) \left( \sigma \times {r}_b \right) ^j \right] \right.
\\&
\left. +\left[ \left( \sigma \times {r}_a \right) ^iG_0\left( \omega \right) r_{b}^{j}+\left( \sigma \times {r}_a \right) ^iG_0\left( -\omega \right) r_{b}^{j} \right] \right\},
\end{aligned}
\end{equation}
\begin{equation}\label{rel_tensor}
\begin{aligned}
\hat{\alpha}_{T,R}^{ij}&
=-e^2\sum_{a,b}{}\left[ H_R\frac{Q}{E_0-H_0}r_{a}^{\{i}\frac{2\left( E_0-H_0 \right)}{\left( E_0-H_0 \right) ^2-\omega ^2}r_{b}^{j\}}+r_{a}^{\{i}\frac{2\left( E_0-H_0 \right)}{\left( E_0-H_0 \right) ^2-\omega ^2}r_{b}^{j\}}\frac{Q}{E_0-H_0}H_R \right]
\\&
-e^2\sum_{a,b}{}\left[ r_{a}^{\{i}G^{\left( 1 \right)}\left( -\omega \right) r_{b}^{j\}}+r_{a}^{\{i}G^{\left( 1 \right)}\left( \omega \right) r_{b}^{j\}} \right] -\hat{\alpha}_{S,R}\,.
\end{aligned}
\end{equation}
Higher-order relativistic corrections can, in principle, be obtained in a similar way using more complicated nonrelativistic expansions and interaction vertices.

\subsection{Leading-Order Radiative Corrections}
In the following, we take a hydrogen-like atom as an example, where the nonrelativistic energy correction Eq.~(\ref{NR-energy}) becomes
\begin{equation}
\begin{aligned}
\delta H_{\rm E,NR}=e^2\mathcal{E}_{\text{ex}}^{i}\mathcal{E}_{\text{ex}}^{j\dag}\left[ r^iG_0\left( -\omega \right) r^j+r^jG_0\left( \omega \right) r^i \right],
\end{aligned}
\end{equation}
and the corresponding polarizability components Eqs.~(\ref{alpha_S}), (\ref{alpha_V}), and (\ref{alpha_T}) become
\begin{equation}
\begin{aligned}
\hat{\alpha}_{\rm S,NR}&=-\frac{4}{3}\delta ^{ij}e^2r^i\frac{\left( E_0-H_0 \right)}{\left( E_0-H_0 \right) ^2-\omega ^2}r^j,
\\
\hat{\alpha}_{\rm V,NR}^{k}&=-\text{2}\imath\epsilon ^{ijk}e^2r^i\frac{\omega}{\left( E_0-H_0 \right) ^2-\omega ^2}r^j,
\\
\hat{\alpha}_{\rm T,NR}^{ij}&=e^2r^{\left\{ i \right.}\frac{-2\left( E_0-H_0 \right)}{\left( E_0-H_0 \right) ^2-\omega ^2}r^{j\}}+\frac{4}{3}\delta ^{ij}e^2{}r^{i}\frac{\left( E_0-H_0 \right)}{\left( E_0-H_0 \right) ^2-\omega ^2}r^{j}.
\end{aligned}
\end{equation}
The radiative corrections consist of the high-energy part ($|\bm k| > K\sim m\alpha$) and the low-energy part ($|\bm k|<K$), according to the momentum of virtual photon~\cite{greiner2013quantum}, where $K$ is the momentum cut-off factor that can be eliminated in the final result.
The energy shift due to the high energy virtual photon can be calculated according to 
\begin{equation}\label{high-energy}
\begin{aligned}
\left< \delta H_{\rm E,QED}^{\rm H} \right> &=\left< H_{\rm QED}\frac{Q}{E_0-H_0}\delta H_{\rm E,NR}+\delta H_{\rm E,NR}\frac{Q}{E_0-H_0}H_{\rm QED} \right>
\\&
+e^2\mathcal{E}_{\text{ex}}^{i}\mathcal{E}_{\text{ex}}^{j\dag}\left< \left[ r^iG_{\rm QED}^{\left( 1 \right)}\left( -\omega \right) r^j+r^jG_{\rm QED}^{\left( 1 \right)}\left( \omega \right) r^i \right] \right> ,
\end{aligned}
\end{equation}
where $H_{\rm QED}$ is the QED Hamiltonian given by
\begin{equation}\label{QED-hamiltonian}
\begin{aligned}
H_{\rm QED}=\frac{\alpha}{3\pi m^2}\left( \ln \frac{m}{2K}+\frac{5}{6}-\frac{3}{8}-\frac{1}{5} \right) \nabla^2 V(r)
+\frac{\alpha}{8\pi m^2}
\left[ \nabla^2 V(r)+2\bm{\sigma} \cdot \nabla V(r) \times \bm{p} \right],
\end{aligned}
\end{equation}
with $V(r)$ being the Coulomb potential between the electron and the nucleus, and
$G^{(1)}_{\rm QED}(\pm \omega)=-G_0(\pm \omega) (E_{\rm QED}-H_{\rm QED}) G_0(\pm \omega)$.

The low-energy part can be derived by inserting the irreducible interaction operators $\Sigma$ into the nonrelativistic corrections, leading to
\begin{equation}\label{low-energy}
\begin{aligned}
\left< \delta H_{\rm E,QED}^{\rm L} \right>&=\left< \Sigma _0\frac{Q}{E_0-H_0}\delta H_{\rm E,NR} \right> +\left< \delta H_{\rm E,NR}\frac{Q}{E_0-H_0}\Sigma _0 \right>
\\&
+e^2\mathcal{E}_{\text{ex}}^{i}\mathcal{E}_{\text{ex}}^{j\dag}\left< \left[ r^iG_0\left( -\omega \right) \Sigma _0G_0\left( -\omega \right) r^j+r^jG_0\left( \omega \right) \Sigma _0G_0\left( \omega \right) r^i \right] \right>
\\&
+\mathcal{E}_{\text{ex}}^{i}\mathcal{E}_{\text{ex}}^{j\dag}\left< \left[ \Sigma _{1}^{i}G_0\left( -\omega \right) \left( -er^j \right) +\Sigma _{1}^{j}G_0\left( \omega \right) \left( -er^i \right) \right] \right>
\\&
+\mathcal{E}_{\text{ex}}^{i}\mathcal{E}_{\text{ex}}^{j\dag}\left< \left[ \left( -er^i \right) G_0\left( -\omega \right) \Sigma _{1}^{j}+\left( -er^j \right) G_0\left( \omega \right) \Sigma _{1}^{i} \right] \right>
\\&
+\mathcal{E}_{\text{ex}}^{i}\mathcal{E}_{\text{ex}}^{j\dag}\left< \left( \Sigma _{2}^{ij}+\Sigma _{2}^{ji} \right) \right>
\\&
+\left< \delta H_{\rm E,NR} \right> \left( \partial _{E_0}\Sigma _0 \right) +\left< \Sigma _0 \right> \left( \partial _{E_0}\delta H_{\rm E,NR} \right).
\end{aligned}
\end{equation}
The irreducible interaction operators $\Sigma$ can be understood as the corrections involving the virtual photon loop. The interaction vertices of virtual photon are contributed by $\mathcal{H}_{\text{at}}$. $\Sigma_0$ is the self-energy virtual photon loop, $\Sigma_1$ is the 1-vertex correction, and $\Sigma_2$ is the 2-vertex correction, listed below
\begin{equation}\label{virtual-photon-loop}
\begin{aligned}
\Sigma _0&=e^2\int_{}^K{}\text{d}^3\widetilde{k}D^{kl}\frac{{p}^k}{m}\left( \frac{1}{E_{\xi}-H_0-\omega '}-\frac{1}{-\omega '} \right) \frac{{p}^l}{m}=\Sigma _{0}^{\ln K}+\tilde{\Sigma}_0,
\\
\Sigma _{1}^{i}&=e^2\int_{}^K{}\text{d}^3\widetilde{k}D^{kl}\frac{{p}^k}{m}\frac{1}{E_{\xi}-H_0-\omega '}{r}^i\frac{1}{E_{\eta}-H_0-\omega '}\frac{{p}^l}{m}=\Sigma _{1}^{i\ln K}+\tilde{\Sigma}_{1}^{i},
\\
\Sigma _{2}^{ij}&=e^4\int_{}^K{}\text{d}^3\widetilde{k}D^{kl}\frac{{p}^k}{m}\frac{1}{E_{\xi}-H_0-\omega '}{r}^i\frac{1}{E_{\eta} -H_0-\omega '}{r}^j\frac{1}{E_{\zeta}-H_0-\omega '}\frac{{p}^l}{m},
\end{aligned}
\end{equation}
where $E_\xi$, $E_\eta$, and $E_\zeta$ represent the energies in different propagation stages, which have different values such as $E_0$, $E_0+\omega$, and $E_0-\omega$. Also in the above, $\mathrm{d}^3 \widetilde{k}\equiv \frac{\mathrm{d}^3 k'}{2\omega' (2\pi)^3}$, and $D^{kl}\equiv \delta^{kl}-\frac{k'^k k'^l}{\omega'^2}$ is the photon propagator in the Coulomb gauge with $\omega'$ being the energy of virtual photon.
Finally $K$ is the energy upper bound to the low-energy virtual photon.
These integrals have ultraviolet divergence when the virtual photon momentum approaches infinity, except for $\Sigma_2$. The linear divergence of $\Sigma_0$ can be eliminated by subtracting the mass-counter-term $\frac{1}{-\omega'}$~\cite{greiner2013quantum}, as displayed above, while the rest ultraviolet divergence should cancel out with the infrared divergence from the high-energy contribution (the terms containing $K$).
The divergent operators $\Sigma_0$ and $\Sigma_1$ can be written as the sum of $\Sigma^{\ln K}$ and $\tilde{\Sigma}$, where $\tilde{\Sigma}$ is the finite part not containing the cut-off factor $K$.
Here we list the results after performing the integration of $\Sigma_n^{\ln K}$, $\tilde{\Sigma}_n$ $(n=0,1)$, and $\Sigma_2$
\begin{equation}\label{counter-operator}
\begin{aligned}
\Sigma _{0}^{\ln K}&=\frac{-2\alpha}{3\pi m^2}{p}^k\left( E_{\xi}-H_0 \right) {p}^k\ln \left( 2K \right) ,
\\
\Sigma _{1}^{i\ln K}&=\frac{2\alpha}{3\pi m^2} {p}^k {r}^i {p}^k\ln \left( 2K \right) ,
\\
\tilde{\Sigma}_0&=\frac{2\alpha}{3\pi m^2} {p}^k\left( E_{\xi}-H_0 \right) {p}^k \ln |2\left( E_{\xi}-H_0 \right) |,
\\
\tilde{\Sigma}_{1}^{i}&=\frac{-2\alpha}{3\pi m^2}\sum_{1,2}
{p}^k\left| 1 \right> \left< 1 \right|{r}^i\left| 2 \right> \left< 2 \right|{p}^k\frac{E_{\eta ,2}\ln |2E_{\eta ,2}|-E_{\xi ,1}\ln |2E_{\xi ,1}|}{E_{\xi ,1}-E_{\eta ,2}},
\end{aligned}
\end{equation}
and
\begin{equation}
\begin{aligned}
\Sigma _{2}^{ij}
=&
\frac{-2\alpha}{3\pi m^2}
\sum_{\text{1,2,}3}{}
{p}^k
\left| 1 \right> \left< \text{1}|
{r}^i|2 \right> \left< \text{2}|
{r}^j|3 \right> \left< 3 \right|{p}^k
\left\{ \frac{E_{\xi ,1}\ln |E_{\xi ,1}|}{\left( E_{\xi ,1}-E_{\eta ,2} \right) \left( E_{\eta ,3}-E_{\xi ,1} \right)} \right.
\\&
+\frac{E_{\eta ,2}\ln |E_{\eta ,2}|}{\left( E_{\eta ,2}-E_{\zeta ,3} \right) \left( E_{\xi ,1}-E_{\eta ,2} \right)}
+\left. \frac{E_{\zeta ,3}\ln |E_{\zeta ,3}|}{\left( E_{\zeta ,3}-E_{\xi ,1} \right) \left( E_{\eta ,2}-E_{\eta ,3} \right)} \right\}\,,
\end{aligned}
\end{equation}
where $E_{s,t}\equiv E_{s}-E_t$, and $E_i=\langle i | H_0 | i \rangle, (i=1,2,3)$, are the energies of the intermediate states $|1\rangle, |2\rangle$, and $|3\rangle$ respectively. The summation over $1,2,3$ means summation over all possible intermediate states $|1\rangle, |2\rangle$ and $|3\rangle$. The result of the $\Sigma_2$ integral does not contain $\ln K$ as it is finite. The $K$-dependent terms in the high-energy part Eq.~(\ref{high-energy}) and the low-energy part Eq.~(\ref{low-energy}) will cancel out exactly with each other after combining, see Appendix~\ref{appendix_note_2}.

Then the QED contributions to the polarizability components can be extracted from the finite results of the high and low energy parts of the QED
Hamiltonian. The obtained finite results are
\begin{equation}
\begin{aligned}
\hat{\alpha}_{\rm S,QED}&=\tilde{H}_{\rm QED}\frac{Q}{E_0-H_0}\hat{\alpha}_{\rm S,NR}+\hat{\alpha}_{\rm S,NR}\frac{Q}{E_0-H_0}\tilde{H}_{\rm QED}
\\&
+\frac{-2\delta ^{ij}}{3}e^2r^i\left[ \tilde{G}_{\rm QED}^{\left( 1 \right)}\left( -\omega \right) +\tilde{G}_{\rm QED}^{\left( 1 \right)}\left( \omega \right) \right] r^j
\\&
+\tilde{\Sigma}_0\frac{Q}{E_0-H_0}\hat{\alpha}_{\rm S,NR}+\hat{\alpha}_{\rm S,NR}\frac{Q}{E_0-H_0}\tilde{\Sigma}_0
\\&
+\frac{-2\delta ^{ij}e^2}{3}\left[ r^iG_0\left( -\omega \right) \tilde{\Sigma}_0G_0\left( -\omega \right) r^j+r^iG_0\left( \omega \right) \tilde{\Sigma}_0G_0\left( \omega \right) r^j \right]
\\&
+\frac{-2\delta ^{ij}}{3}\left[ \tilde{\Sigma}_{1}^{i}G_0\left( -\omega \right) \left( -er^j \right) +\tilde{\Sigma}_{1}^{i}G_0\left( \omega \right) \left( -er^j \right) \right]
\\&
+\frac{-2\delta ^{ij}}{3}\left[ \left( -er^i \right) G_0\left( -\omega \right) \tilde{\Sigma}_{1}^{j}+\left( -er^i \right) G_0\left( \omega \right) \tilde{\Sigma}_{1}^{j} \right]
\\&
+\frac{-2\delta ^{ij}}{3}\left( \Sigma _{2}^{ij}+\Sigma _{2}^{ji} \right),
\end{aligned}
\end{equation}
\begin{equation}
\begin{aligned}
\hat{\alpha}_{\rm V,QED}^{k}&
=\tilde{H}_{\rm QED}\frac{Q}{E_0-H_0}\hat{\alpha}_{\rm V,NR}^{k}+\hat{\alpha}_{\rm V,NR}^{k}\frac{Q}{E_0-H_0}\tilde{H}_{\rm QED}
\\&
-\imath\epsilon ^{ijk}e^2r^i\left[ \tilde{G}_{\rm QED}^{\left( 1 \right)}\left( -\omega \right) -\tilde{G}_{\rm QED}^{\left( 1 \right)}\left( \omega \right) \right] r^j
\\&
+\tilde{\Sigma}_0\frac{Q}{E_0-H_0}\hat{\alpha}_{\rm V,NR}^{k}+\hat{\alpha}_{\rm V,NR}^{k}\frac{Q}{E_0-H_0}\tilde{\Sigma}_0
\\&
-\imath\epsilon ^{ijk}e^2\left[ r^iG_0\left( -\omega \right) \tilde{\Sigma}_0G_0\left( -\omega \right) r^j-r^iG_0\left( \omega \right) \tilde{\Sigma}_0G_0\left( \omega \right) r^j \right]
\\&
-\imath\epsilon ^{ijk}\left[ \tilde{\Sigma}_{1}^{i}G_0\left( -\omega \right) \left( -er^j \right) -\tilde{\Sigma}_{1}^{i}G_0\left( \omega \right) \left( -er^j \right) \right]
\\&
-\imath\epsilon ^{ijk}\left[ \left( -er^i \right) G_0\left( -\omega \right) \tilde{\Sigma}_{1}^{j}-\left( -er^i \right) G_0\left( \omega \right) \tilde{\Sigma}_{1}^{j} \right]
\\&
-\imath\epsilon ^{ijk}\left( \Sigma _{2}^{ij}-\Sigma _{2}^{ji} \right) ,
\end{aligned}
\end{equation}
\begin{equation}
\begin{aligned}
\hat{\alpha}_{\rm T,QED}^{ij}&=
\tilde{H}_{\rm QED}\frac{Q}{E_0-H_0}\hat{\alpha}_{\rm T,NR}^{ij}+\hat{\alpha}_{\rm T,NR}^{ij}\frac{Q}{E_0-H_0}\tilde{H}_{\rm QED}
\\&
+e^2r_{a}^{\{i}\left[ \tilde{G}_{\rm QED}^{\left( 1 \right)}\left( -\omega \right) +\tilde{G}_{\rm QED}^{\left( 1 \right)}\left( \omega \right) \right] r_{b}^{j\}}
\\&
+\tilde{\Sigma}_0\frac{Q}{E_0-H_0}\hat{\alpha}_{\rm T,NR}^{ij}+\hat{\alpha}_{\rm T,NR}^{ij}\frac{Q}{E_0-H_0}\tilde{\Sigma}_0
\\&
+e^2\left[ r^{\{i}G_0\left( -\omega \right) \tilde{\Sigma}_0G_0\left( -\omega \right) r^{j\}}+r^{\{i}G_0\left( \omega \right) \tilde{\Sigma}_0G_0\left( \omega \right) r^{j\}} \right]
\\&
-\left[ \tilde{\Sigma}_{1}^{\{i}G_0\left( -\omega \right) \left( -er^{j\}} \right) +\tilde{\Sigma}_{1}^{\{j}G_0\left( \omega \right) \left( -er^{i\}} \right) \right]
\\&
-\left[ \left( -er^{\{i} \right) G_0\left( -\omega \right) \tilde{\Sigma}_{1}^{j\}}+\left( -er^{\{j} \right) G_0\left( \omega \right) \tilde{\Sigma}_{1}^{i\}} \right]
\\&
+\Sigma _{2}^{\left\{ ij \right\}}+\Sigma _{2}^{\left\{ ji \right\}}-\hat{\alpha}_{\rm S,QED}\,,
\end{aligned}
\end{equation}
where $\tilde{H}_{\rm QED}$ and $\tilde{G}^{(1)}_{\rm QED}$ are, respectively, the remainders after cancelling out the $K$-containing terms in $H_{\rm QED}$ and $G^{(1)}_{QED}$.

The Bethe-logarithm correction, denoted as $\ln k_0$, shows up in the low-energy treatment, corresponding to the operator $\tilde{\Sigma}_0$. The other two operators $\tilde{\Sigma}_1$ and $\Sigma_2$ contribute to other types of logarithmic terms. The numerical evaluation of the second-order electric-field derivative of the Bethe-logarithm is very difficult~\cite{PhysRevLett.92.233001}.
However, the integral representation of $\ln k_0$ can be used instead
\begin{equation}\label{integral lnk}
\begin{aligned}
\ln k_{0}=\lim_{\Lambda \rightarrow \infty}
\left[-\frac{\left< p^{2} \right>}{\mathcal{D}}
\Lambda+\ln(2\Lambda)
+
\int_{0}^{\Lambda}\omega\frac{J(\omega)}{\mathcal{D}}d\omega\right],
\end{aligned}
\end{equation}
where $\mathcal{D}=4\pi\langle \delta^3 (\boldsymbol{r})\rangle$.
The second-order electric field derivative of $J(\omega)$ is given in Ref.~\cite{PhysRevLett.92.233001}
\begin{equation}\label{2nd J}
\begin{aligned}
\partial _{\mathcal{E}}^{2}J\left( \omega \right)
&=
\frac{2}{3}\left[ 2\left< \psi _0 \right|r^i\mathcal{R}_0r^i\mathcal{R}_0p^j\mathcal{R}\left( \omega \right) p^j\left| \psi _0 \right> \right. +2\left< \psi _0 \right|r^i\mathcal{R}_0p^j\mathcal{R}\left( \omega \right) r^i\mathcal{R}_0p^j\left| \psi _0 \right>
\\&
+\left< \psi _0 \right|r^i\mathcal{R}_0p^j\mathcal{R}\left( \omega \right) p^j\mathcal{R}_0r^i\left| \psi _0 \right> -\left< \psi _0\left| p^j\mathcal{R}\left( \omega \right) p^j \right|\psi _0 \right> \left< \psi _0\left| r^i\mathcal{R}_{0}^{2}r^i \right|\psi _0 \right>
\\&
\left. +\left< \psi _0\left| p^j\mathcal{R}\left( \omega \right) r^i\mathcal{R}\left( \omega \right) r^i\mathcal{R}\left( \omega \right) p^j \right|\psi _0 \right> -\left< \psi _0\left| r^i\mathcal{R}_0r^i \right|\psi _0 \right> \left< \psi _0\left| p^j\mathcal{R}\left( \omega \right) ^2p^j \right|\psi _0 \right> \right],
\end{aligned}
\end{equation}
where $\omega$ means the energy of virtual photon, which is $\omega'$ in our case. For the case of static field where the frequency of external field is zero, $\mathcal{R}_0 = Q(E_0-H_0)^{-1}$, instead of $G_0(\omega)$. Except for these differences, \eqref{integral lnk} and \eqref{2nd J} are consistent with our calculations of the low-energy parts. In our calculations, however, whether the Bethe logarithm is independent of electric field or not is treated in a unified scheme. Additionally, we can also obtain higher-order relativistic interaction vertices as shown in Appendix~\ref{appendix_note_2}.

\subsection{Static Limit}
By applying the static limit $\omega\rightarrow 0$, we can compare our relativistic correction with the known result
of the static polarizability~\cite{PhysRevA.63.012504}. The relativistic corrections to the dynamical polarizability are given
in Eqs.~\eqref{rel_scalar}, \eqref{rel_vector}, and \eqref{rel_tensor}. In our scheme, by making the external electric field to be real-valued, the energy shift under zero $\omega$ becomes
\begin{equation}
\begin{aligned}
\delta H_{\rm d}= e^2 \sum_{a,b} r^i_a\mathcal{E}^i_{\text{ex}} \frac{1}{E_0-H_0} r^j_b \mathcal{E}^j_{\text{ex}}\,.
\end{aligned}
\end{equation}
For a static field with the energy shift being in the form
\begin{equation}
\begin{aligned}
\Delta E \sim \frac{-1}{2} \alpha_{\rm d} \mathcal{E}^2_{\text{ex}},
\end{aligned}
\end{equation}
one can evaluate it using the following part of $\delta H_{\rm d}$
\begin{equation}
\begin{aligned}
\delta H_{\rm d}\sim \frac{e^2}{3} \delta^{ij}\mathcal{E}^i_{\text{ex}} \mathcal{E}^j_{\text{ex}}
\sum_{a,b} \bm{r}_a \frac{1}{E_0-H_0} \bm{r}_b\,.
\end{aligned}
\end{equation}
The extracted polarizability is thus
\begin{equation}
\begin{aligned}
\alpha_{\rm d} = \frac{-2}{3} \sum_{a,b} \bm{r}_a \frac{1}{E_0-H_0} \bm{r}_b,
\end{aligned}
\end{equation}
which is the same as Eq.~(3) of Ref.~\cite{PhysRevA.63.012504}.
The $\omega$-dependent relativistic correction is given by Eq.~\eqref{rel_correction},
where the term $\frac{-\imath}{8m^2}\left[ \bm{\sigma}\cdot \bm{\pi},\bm{\sigma}\cdot \bm{\mathcal{E}}_{\text{ex}} \right]$ would vanish under the static limit. Since the two terms involving $G^{(1)}(\pm \omega)$ in Eq.~\eqref{rel_correction} are equal for the static case, a factor of $1/2$ should be multiplied. Replacing $\delta H_{\rm E,NR}$ by $\delta H_{\rm d}$ and making the external field $\mathcal{E}$ real-valued, we have
\begin{equation}
\begin{aligned}
\delta H_{\rm E,R}\sim \delta H_{\rm d,R}
&\equiv
H_{\rm R}\frac{Q}{E_0-H_0}\delta H_{\rm d}+\delta H_{\rm d}\frac{Q}{E_0-H_0}H_{\rm R}
\\&
+
\frac{-2e^2}{3} \delta^{ij} \mathcal{E}^i_{\text{ex}} \mathcal{E}^j_{\text{ex}}
\sum_{a,b} \bm{r}_a \frac{1}{E_0-H_0} (E_{\rm R}-H_{\rm R}) \frac{1}{E_0-H_0} \bm{r}_b\,,
\end{aligned}
\end{equation}
which is equivalent to the relativistic expression in Ref.~\cite{PhysRevA.63.012504}.
Finally, the terms involving $\left[ \bm{\sigma}\cdot \bm{\pi},\bm{\sigma}\cdot \bm{\mathcal{E}}_{\text{ex}} \right]$
in Eq.~\eqref{rel_correction} gives rise to the relativistic corrections to the dynamical polarizability, which
is still less investigated.

\section{Magnetic Polarizability} \label{magnetic polarizability}
When we consider an atom in an external magnetic field, atomic energy levels will be shifted according to the Zeeman effect.
The relativistic and radiative corrections to the Zeeman effect are studied in \cite{PhysRevA.69.052502}.
Here we will consider the Zeeman effect up to the second order in external magnetic field, which can be described by a quantity called the
the magnetic polarizability. We start with the
Hamiltonian $\mathcal{H}_{\rm M}$ expressed in Eq.~(\ref{H_magnetic_hami}).
Usually, the magnetic dipole effect is a factor of $\alpha$ smaller than the electric dipole effect. With this in mind, we obtain
the following Hamiltonian in its leading order
\begin{equation}\label{magnetic_hami_sim}
\begin{aligned}
\mathcal{H}_{\rm M}&\simeq
\frac{-e}{2m}\bm{\mathcal{B}}_{\text{ex}}\cdot \left( \bm{L}+\bm{\sigma} \right) +\frac{-e}{4m}\bm{A}_{\text{at}} \cdot \left( \bm{r}\times \bm{\mathcal{B}}_{\text{ex}} \right)
+\frac{-e}{8m^2}\left( \bm{\sigma} \times \bm{\mathcal{E}}_{\text{at}} \right) \cdot \left( \bm{r}\times \bm{\mathcal{B}}_{\text{ex}} \right)
\\&
+\frac{e}{8m^3}\left\{ \bm{p}^2,\bm{\mathcal{B}}_{\text{ex}}\cdot \left( \bm{L}+\bm{\sigma} \right) \right\}
+\frac{e^2}{8m}\left( \bm{r}\times \bm{\mathcal{B}}_{\text{ex}} \right) ^2+\frac{-e^2}{32m^3}\left\{ \bm{p}^2,\left( \bm{r}\times \bm{\mathcal{B}}_{\text{ex}} \right) ^2 \right\}
\\&
+\frac{-e^2}{8m^3}\left[ \bm{\mathcal{B}}_{\text{ex}}\cdot \left( \bm{L}+\bm{\sigma} \right) \right] ^2\,,
\end{aligned}
\end{equation}
which includes single-photon and double-photon interaction.

\subsection{Nonrelativistic Magnetic Polarizability}
According to Eq.~\eqref{magnetic_hami_sim}, the nonrelativistic interaction is described by
$$
\frac{-e}{2m}\bm{\mathcal{B}}_{\text{ex}}\cdot \left( \bm{L}+\bm{\sigma} \right) +\frac{e^2}{8m}\left( \bm{r}\times \bm{\mathcal{B}}_{\text{ex}} \right) ^2,
$$
where the first term describes the single-photon interaction and the second term the double-photon interaction. The energy shift caused by this interaction is
\begin{equation}
\begin{aligned}
\delta H_{\rm M,NR}&=
\frac{e^2}{4m^2}\sum_{a,b}{}\bm{\mathcal{B}}_{\text{ex}}\cdot \left( \bm{L}+\bm{\sigma} \right) _a\frac{1}{E_0-H_0-\omega}\left( \bm{L}+\bm{\sigma} \right) _b\cdot \bm{\mathcal{B}}_{\text{ex}}^{\dag}
\\&
+\frac{e^2}{4m^2}\sum_{a,b}{}\bm{\mathcal{B}}_{\text{ex}}^{\dag}\cdot \left( \bm{L}+\bm{\sigma} \right) _a\frac{1}{E_0-H_0+\omega}\left( \bm{L}+\bm{\sigma} \right) _b\cdot \bm{\mathcal{B}}_{\text{ex}}
\\&
+\sum_a{}\frac{e^2}{4m}\left( \bm{r}\times \bm{\mathcal{B}}_{\text{ex}} \right) _{a}^{2}\,.
\end{aligned}
\end{equation}
We then divide $\delta H_{\rm M,NR}$ into symmetric and anti-symmetric parts rather than scalar, vector, and tensor parts
\begin{equation}
\begin{aligned}
\delta
H_{\rm M,NR}^{\rm S}&=\frac{e^2}{4m^2}\sum_{a,b}{}\frac{\mathcal{B}_{\text{ex}}^{i}\mathcal{B}_{\text{ex}}^{j\dag}+\mathcal{B}_{\text{ex}}^{j}\mathcal{B}_{\text{ex}}^{i\dag}}{2}\left( L+\sigma \right) _{a}^{i}\frac{2\left( E_0-H_0 \right)}{\left( E_0-H_0 \right) ^2-\omega ^2}\left( L+\sigma \right) _{b}^{j}
\\&
+\frac{e^2}{4m}\sum_a{}\frac{\mathcal{B}_{\text{ex}}^{i}\mathcal{B}_{\text{ex}}^{j\dag}+\mathcal{B}_{\text{ex}}^{i\dag}\mathcal{B}_{\text{ex}}^{j}}{2}\left( \boldsymbol{r}_a^2\delta ^{ij}-r_a^i r_a^j \right) ,
\\
\delta H_{\rm M,NR}^{\rm A}&=\frac{e^2}{4m^2}\sum_{a,b}{}\frac{\mathcal{B}_{\text{ex}}^{i}\mathcal{B}_{\text{ex}}^{j\dag}-\mathcal{B}_{\text{ex}}^{j}\mathcal{B}_{\text{ex}}^{i\dag}}{2}\left\{ \left( L+\sigma \right) _{a}^{i}\frac{2\omega}{\left( E_0-H_0 \right) ^2-\omega ^2}\left( L+\sigma \right) _{b}^{j} \right\},
\end{aligned}
\end{equation}
where the second term in $\delta H^{\rm S}_{\rm M,NR}$ is the double-photon contribution. And the rest terms contribute to the magnetic dipole correction. The second-order energy shift due to the magnetic field can be written in the form
\begin{equation}
\begin{aligned}
\Delta E\sim \frac{-1}{2} \beta^{ij} \mathcal{B}^{i\dag}_{\text{ex}}\mathcal{B}^{j}_{\text{ex}}\,,
\end{aligned}
\end{equation}
where the nonrelativistic magnetic polarizability components can then be extracted
\begin{equation}
\begin{aligned}
\beta^{{\rm S},ij}_{\rm NR}&=
\frac{-e^2}{4m^2}\sum_{a,b}{}\left( L+\sigma \right) _{a}^{\{i}\frac{2\left( E_0-H_0 \right)}{\left( E_0-H_0 \right) ^2-\omega ^2}\left( L+\sigma \right) _{b}^{j\}}
+\frac{-e^2}{4m}\sum_a{}
\left( \bm{r}_a^2\delta ^{\{ij\}}-r_a^{\{i}r_a^{j\}} \right)
,
\\
\beta_{\rm NR}^{{\rm A},k}
&=\frac{-e^2}{4m^2}\epsilon ^{ijk}\sum_{a,b}{}\left\{ \left( L+\sigma \right) _{a}^{i}\frac{2\omega}{\left( E_0-H_0 \right) ^2-\omega ^2}\left( L+\sigma \right) _{b}^{j} \right\}.
\end{aligned}
\end{equation}

\subsection{Relativistic Corrections to the Magnetic Polarizability}
The leading-order relativistic magnetic Hamiltonian is
\begin{equation}
\begin{aligned}
H_{\rm M,R}&=\frac{-e}{4m}\bm{A}_{\text{at}}\cdot \left( \bm{r}\times \bm{\mathcal{B}}_{\text{ex}} \right) +\frac{e}{8m^3}\left\{ \bm{p}^2,\bm{\mathcal{B}}_{\text{ex}}\cdot \left( \bm{L}+\bm{\sigma} \right) \right\} +\frac{-e}{8m^2}\left( \bm{\sigma} \times \bm{\mathcal{E}_{\text{at}}} \right) \cdot \left( \bm{r}\times \bm{\mathcal{B}}_{\text{ex}} \right)
\\&
+\frac{-e^2}{32m^3}\left\{ \bm{p}^2,\left( \bm{r}\times \bm{\mathcal{B}}_{\text{ex}} \right) ^2 \right\} +\frac{-e^2}{8m^3}\left[ \bm{\mathcal{B}}_{\text{ex}}\cdot \left( \bm{L}+\bm{\sigma} \right) \right] ^2,
\end{aligned}
\end{equation}
which contains the kinetic correction of nonrelativistic interaction, the atomic-magnetic coupling field, and the double-photon correction of magnetic-angular momentum interaction. Since the expression of the energy shift is tedious to write down, we only list the magnetic polarizability
components
\begin{equation}
\begin{aligned}
\beta _{\rm R}^{{\rm S},ij}&=
H_{\rm R}\frac{Q}{E_0-H_0} \beta^{{\rm S},ij}_{\rm NR}
+\beta^{{\rm S},ij}_{\rm NR} \frac{Q}{E_0-H_0}H_{\rm R}
\\&
+\frac{-e^2}{4m^2}\sum_{a,b}{}\left( L+\sigma \right) _{a}^{\{i}\left[ G^{\left( 1 \right)}\left( -\omega \right) +G^{\left( 1 \right)}\left( \omega \right) \right] \left( L+\sigma \right) _{b}^{j\}}
\\&
+\frac{-e}{2m}\sum_{a,b}{}\mathcal{J}_{a}^{\{i}\frac{-2\left( E_0-H_0 \right)}{\left( E_0-H_0 \right) ^2-\omega ^2}\left( L+\sigma \right) _{b}^{j\}}
\\&
+\frac{-e}{2m}\sum_{a,b}{}\left( L+\sigma \right) _{a}^{\{i}\frac{-2\left( E_0-H_0 \right)}{\left( E_0-H_0 \right) ^2-\omega ^2}\mathcal{J}_{b}^{j\}}
\\&
+\frac{e^2}{32m^3}\sum_a \left\{ \hat{p}^2,\left( \delta ^{\{ij\}}\bm{r}_{a}^{2}-r_{a}^{\{i}r_{a}^{j\}} \right) \right\} 
+\frac{e^2}{8m^3}\sum_a \left( L+\sigma \right)_a ^{\{i}\left( L+\sigma \right)_a ^{j\}},
\end{aligned}
\end{equation}

\begin{equation}
\begin{aligned}
\beta _{\rm R}^{{\rm A},k}&=H_{\rm R}\frac{Q}{E_0-H_0} \beta^{{\rm A},k}_{\rm NR}
+\beta^{{\rm A},k}_{\rm NR} \frac{Q}{E_0-H_0}H_{\rm R}
\\&
+\frac{-e^2}{4m^2}\epsilon ^{ijk}\sum_{a,b}{}\left( L+\sigma \right) _{a}^{i}\left[ G^{\left( 1 \right)}\left( -\omega \right) -G^{\left( 1 \right)}\left( \omega \right) \right] \left( L+\sigma \right) _{b}^{j}
\\&
+\sum_{a,b}{}\epsilon ^{ijk}\mathcal{J}_{a}^{i}\frac{-2\omega}{\left( E_0-H_0 \right) ^2-\omega ^2}\frac{-e}{2m}\left( L+\sigma \right) _{b}^{j}
\\&
+\sum_{a,b}{}\epsilon ^{ijk}\frac{-e}{2m}\left( L+\sigma \right) _{a}^{i}\frac{-2\omega}{\left( E_0-H_0 \right) ^2-\omega ^2}\mathcal{J}_{b}^{j}\,,
\end{aligned}
\end{equation}
where $G^{(1)}(\pm \omega)$ is defined by Eq.~(\ref{green_G1}) and
\begin{equation}\label{current_operator}
\begin{aligned}
\mathcal{J}_{a}^{i}
=
\frac{-e}{4m}&\left( \bm{A}_{\text{at}} \times \bm{r}_a \right) ^i+\frac{e}{8m^3}\left\{ \bm{p}^2,\left( {L}+{\sigma} \right) ^i \right\} _a+\frac{-e}{8m^2}\left( \bm{\sigma} \times \bm{\mathcal{E}_{\text{at}}}\times \bm{r} \right) _{a}^{i}.
&
\end{aligned}
\end{equation}

\section{Other types of Corrections} \label{other types of corrections}
Other types of corrections can also be obtained by recalling the Hamiltonians in Sec.~\ref{theory and method}, including
external electric-magnetic coupling interaction, or the Coulomb-transverse photon vertex. Multipole interactions are contained in the PZ transformed vector potential, as indicated in Eqs.~\eqref{multipole_indi} and \eqref{electric_hami}. The magnetic multipole terms are discussed in
Appendix~\ref{appendix_note_1}. Here we first discuss the external electric-magnetic coupling or Coulomb-transverse photon contributions by considering the leading-order relativistic corrections, followed by a brief discussion on some types of multipole corrections.

\subsection{Coulomb-Transverse Photon Contribution}
Consider the electric dipole interaction $-er^i\mathcal{E}_{\text{ex}}^i$, the magnetic dipole interaction $\frac{-e}{2m}\bm{\mathcal{B}}_{\text{ex}}\cdot \left( \bm{L}+\bm{\sigma} \right)$, and the double-photon interaction $\mathcal{H}_{EM}$ defined in Eq.~(\ref{H_electric_magnetic_hami}). The nonrelativistic contribution of the coupling among these interactions is
\begin{eqnarray}
\delta H_{\rm EM,NR}&=&
\frac{e^2}{2m}\mathcal{E}_{\text{ex}}^{i} \mathcal{B}_{\text{ex}}^{j\dag}\sum_{a,b}{}\left[ r_{a}^{i}\frac{1}{E_0-H_0-\omega}\left( L+\sigma \right) _{b}^{j}+\left( L+\sigma \right) _{a}^{j}\frac{1}{E_0-H_0+\omega}r_{a}^{i} \right]
\nonumber\\
&+&\frac{e^2}{2m}\mathcal{E}_{\text{ex}}^{i\dag} \mathcal{B}_{\text{ex}}^{j}\sum_{a,b}{}\left[ r_{a}^{i}\frac{1}{E_0-H_0+\omega}\left( L+\sigma \right) _{b}^{j}+\left( L+\sigma \right) _{a}^{j}\frac{1}{E_0-H_0-\omega}r_{a}^{i} \right]
\nonumber\\
&+&\frac{-e}{8m^2}\sum_a{}\frac{1}{2}
\left[ \left( \bm{\sigma}_a \times \bm{\mathcal{E}}_{\text{ex}}^{\dag} \right) \cdot \left( \bm{r}_a \times \bm{\mathcal{B}}_{\text{ex}} \right)
+\left( \bm{\sigma}_a \times \bm{\mathcal{E}}_{\text{ex}} \right) \cdot \left( \bm{r}_a \times \bm{\mathcal{B}}_{\text{ex}}^{\dag} \right) \right].
\end{eqnarray}
It is seen that the contributions from the first two terms above are zero, because the electric and magnetic dipole operators have opposite
parities. The last term above can be simplified as
\begin{equation}
\begin{aligned}
\frac{-e}{8m^2}\frac{\mathcal{B}_{\text{ex}}^{i}\mathcal{E}_{\text{ex}}^{j\dag}+\mathcal{B}_{\text{ex}}^{i\dag}\mathcal{E}_{\text{ex}}^{j}}{2}
\sum_a{}\left( \delta ^{ij}\left( \bm{\sigma}\cdot \bm{r} \right) -\sigma ^ir^j \right) _a\,,
\end{aligned}
\end{equation}
which also has zero contribution, due to $\langle \bm{r} \rangle=0$ for a nonrelativistic wave function of fixed parity.

Now we consider higher-order corrections to see if there are non-zero contributions.
The leading-order relativistic correction to the electric dipole is generated by the relativistic electric-dipole interaction with the nonrelativistic magnetic dipole interaction, and by the relativistic magnetic dipole interaction with the nonrelativistic electric interaction, shown below
\begin{equation}
\begin{aligned}
\Delta H_{\rm EM,R}&=H_{\rm R}\frac{Q}{E-H_0}\Delta H_{\rm EM,NR}+\Delta H_{\rm EM,NR}\frac{Q}{E-H_0}H_{\rm R}
\\&
+\frac{e^2}{2m}\mathcal{E}_{\text{ex}}^{i}\mathcal{B}_{\text{ex}}^{j\dag}\sum_{a,b}{}\left[ r_{a}^{i}G^{\left( 1 \right)}\left( -\omega \right) \left( L+\sigma \right) _{b}^{j}+\left( L+\sigma \right) _{a}^{j}G^{\left( 1 \right)}\left( \omega \right) r_{a}^{i} \right]
\\&
+\frac{e^2}{2m}\mathcal{E}_{\text{ex}}^{i\dag}\mathcal{B}_{\text{ex}}^{j}\sum_{a,b}{}\left[ r_{a}^{i}G^{\left( 1 \right)}\left( \omega \right) \left( L+\sigma \right) _{b}^{j}+\left( L+\sigma \right) _{a}^{j}G^{\left( 1 \right)}\left( -\omega \right) r_{a}^{i} \right]
\\&
-e\mathcal{E}_{\text{ex}}^{i}\mathcal{B}_{\text{ex}}^{j\dag}\sum_{a,b}{}\left[ r_{a}^{i}\frac{1}{E_0-H_0-\omega}\mathcal{J}_{b}^{j}+\mathcal{J}_{a}^{j}\frac{1}{E_0-H_0+\omega}r_{b}^{i} \right]
\\&
-e\mathcal{E}_{\text{ex}}^{i\dag}\mathcal{B}_{\text{ex}}^{j}\sum_{a,b}{}\left[ r_{a}^{i}\frac{1}{E_0-H_0+\omega}\mathcal{J}_{b}^{j}+\mathcal{J}_{a}^{j}\frac{1}{E_0-H_0-\omega}r_{b}^{i} \right]
\\&
+\frac{-\imath e}{8m^2}\mathcal{E}_{\text{ex}}^{i}\mathcal{B}_{\text{ex}}^{j\dag}\sum_a{}\left[ \left( \boldsymbol{\sigma }\times \boldsymbol{r} \right) _{a}^{i},\left( L+\sigma \right) _{a}^{j} \right] +\frac{-ie}{8m^2}\mathcal{E}_{\text{ex}}^{i\dag}\mathcal{B}_{\text{ex}}^{j}\sum_a{}\left[ \left( \bm{\sigma} \times \boldsymbol{r} \right) _{a}^{i},\left( L+\sigma \right) _{a}^{j} \right]
\\&
+\frac{-\imath e}{8m^2}\mathcal{E}_{\text{ex}}^{i}\mathcal{B}_{\text{ex}}^{j\dag}\sum_{a,b}{}\left[ \left( \boldsymbol{\sigma }\times \boldsymbol{r} \right) _{a}^{i}\frac{\omega}{E-H_0-\omega}\left( L+\sigma \right) _{b}^{j}+\left( L+\sigma \right) _{a}^{j}\frac{\omega}{E-H_0+\omega}\left( \boldsymbol{\sigma }\times \boldsymbol{r} \right) _{b}^{i} \right]
\\&
+\frac{-\imath e}{8m^2}\mathcal{E}_{\text{ex}}^{i\dag}\mathcal{B}_{\text{ex}}^{j}\sum_{a,b}{}\left[ \left( \boldsymbol{\sigma }\times \boldsymbol{r} \right) _{a}^{i}\frac{-\omega}{E-H_0+\omega}\left( L+\sigma \right) _{b}^{j}+\left( L+\sigma \right) _{a}^{i}\frac{-\omega}{E-H_0-\omega}\left( \boldsymbol{\sigma }\times \boldsymbol{r} \right) _{b}^{j} \right],
\end{aligned}
\end{equation}
where the current operator is defined as Eq.~\eqref{current_operator} and $G^{(1)}(\pm \omega)$ defined as Eq.~\eqref{green_G1}.

The contribution from this higher-order interaction is also zero, because the electric and magnetic vertices have opposite parity and the Hamiltonian $H_0$ in the propagator is a parity-even operator. Furthermore, there is no way for the intermediate states to break the parity, no matter how many they have. Therefore, this type of interaction should have no contribution to the polarizabilities. However, if $H_0$ contains some non-perturbative terms that break the parity symmetry, there may exist non-zero contributions to the polarizabilities.

\subsection{Multipole Contributions}
In principle, the multipole interaction of arbitrary order can be extracted from our scheme. The
electric multipole terms are contained in Eq.~\eqref{A multipole form} and the magnetic multipole terms can be seen
in Appendix~\ref{appendix_note_1}. We will not consider the magnetic multipole interactions as they are higher order of magnitude and thus negligible. Here we present the energy shifts due to the electric dipole-quadrupole, electric dipole-octupole, and electric quadrupole-magnetic dipole couplings. The electric dipole-quadrupole interaction term is
\begin{equation}
\begin{aligned}
\delta H_{\rm E1E2}
&=\frac{e^2}{2}\sum_{a,b}{}\left[ r_{a}^{i}\mathcal{E}_{\text{ex}}^{i}\frac{1}{E_0-H_0-\omega}r_{b}^{j}r_{b}^{k}\left( \nabla ^k\mathcal{E}_{\text{ex}}^{j\dag} \right) +r_{a}^{i}\mathcal{E}_{\text{ex}}^{i\dag}\frac{1}{E_0-H_0+\omega}r_{b}^{j}r_{b}^{k}\left( \nabla ^k\mathcal{E}_{\text{ex}}^{j} \right) \right]
\\&
+\frac{e^2}{2}\sum_{a,b}{}\left[ r_{a}^{i}r_{a}^{k}\left( \nabla ^k\mathcal{E}_{\text{ex}}^{i} \right) \frac{1}{E_0-H_0-\omega}r_{b}^{j}\mathcal{E}_{\text{ex}}^{j\dag}+r_{a}^{i}r_{a}^{k}\left( \nabla ^k\mathcal{E}_{\text{ex}}^{i} \right) \frac{1}{E_0-H_0+\omega}r_{b}^{j}\mathcal{E}_{\text{ex}}^{j} \right]\,,
\end{aligned}
\end{equation}
which has zero contribution, because the electric dipole $r^i \mathcal{E}^i_{\text{ex}}$ has odd parity and the electric quadrupole interaction $r^i r^j \nabla^j \mathcal{E}^i$ has even parity.
Notice that the electric dipole and octupole terms have the same parity, which may result in a non-zero contribution. The electric dipole-octupole correction is given by
\begin{equation}
\begin{aligned}
\delta H_{\rm E1E3}
&=
\mathcal{E}_{\text{ex}}^{i}\left( \nabla ^l\mathcal{E}_{\text{ex}}^{j\dag} \right) \frac{e^2}{\text{3!}}\sum_{a,b}{}\left[ r_{a}^{i}\frac{1}{E_0-H_0-\omega}r_{b}^{j}r_{b}^{k}r_{b}^{l}+r_{a}^{j}r_{a}^{k}r_{a}^{l}\frac{1}{E_0-H_0+\omega}r_{b}^{i} \right]
\\&
+\mathcal{E}_{\text{ex}}^{i\dag}\left( \nabla ^l\mathcal{E}_{\text{ex}}^{j} \right) \frac{e^2}{\text{3!}}\sum_{a,b}{}\left[ r_{a}^{i}\frac{1}{E_0-H_0+\omega}r_{b}^{j}r_{b}^{k}r_{b}^{l}+r_{a}^{j}r_{a}^{k}r_{a}^{l}\frac{1}{E_0-H_0-\omega}r_{b}^{i} \right].
\end{aligned}
\end{equation}
Similarly, the electric quadrupole and magnetic dipole terms have the same parity and their coupling gives rise to the following Hamiltonian
\begin{equation}
\begin{aligned}
\delta H_{\rm E2M1}&
=\frac{e^2}{2m}\left( \nabla ^k\mathcal{E}_{\text{ex}}^{i} \right) \mathcal{B}_{\text{ex}}^{j\dag}\sum_{a,b}{}\left[ r_{a}^{i}r_{a}^{k}\frac{1}{E_0-H_0-\omega}\left( L+\sigma \right) _{b}^{j}+\left( L+\sigma \right) _{a}^{j}\frac{1}{E_0-H_0-\omega}r_{b}^{i}r_{b}^{k} \right]
\\&
+\frac{e^2}{2m}\left( \nabla ^k\mathcal{E}_{\text{ex}}^{i\dag} \right) \mathcal{B}_{\text{ex}}^{j}\sum_{a,b}{}\left[ r_{a}^{i}r_{a}^{k}\frac{1}{E_0-H_0+\omega}\left( L+\sigma \right) _{b}^{j}+\left( L+\sigma \right) _{a}^{j}\frac{1}{E_0-H_0-\omega}r_{b}^{i}r_{b}^{k} \right].
\end{aligned}
\end{equation}

\section{Summary} \label{summary}
In this paper, we developed the theory of long-wavelength quantum electrodynamics to describe the second order external electromagnetic effects on light atomic systems, where the relativistic and radiative corrections were treated in a unified scheme.
The nonrelativistic approximation was realized using the FW transformation.
The interaction from external electromagnetic field was distinguished from the atomic interaction by applying the Power-Zienau transformation.
We can see that successive use of the FW and PZ transformations on the Dirac Hamiltonian allows us to obtain the interaction between an atom and a long-wavelength external field, which includes not only the effect of an external field on the atom, but also the relativistic effects inside the atom, as well as the coupling between the two. The effects due to electric and magnetic fields can be clearly separated, and their multipole terms can be identified during the transformation process. Our approach is somewhat different from Pachucki's approach~\cite{PhysRevA.69.052502}, where Pachucki uses the FW Hamiltonian first to construct the nonrelativistic interaction for the atom, {\it i.e.}, the Hamiltonian in quantum mechanics that can be used directly in calculating matrix elements, followed by an application of the PZ transformation to describe the effects brought in by the external fields. In our approach, we first apply the PZ transformation on the FW Hamiltonian, followed by the construction of the interaction from the PZ Hamiltonian, where this PZ Hamiltonian is, in fact, the Hamiltonian density in quantum field theory. However, these two approaches should be equivalent in nonrelativistic quantum field theory.

For the radiative corrections, we treated them at high energy and low energy regions. A cut-off $K$ was introduced to deal with the ultraviolet divergence caused by the virtual photon loop. The similarities and differences of our approach with other work were discussed. We can also obtain higher-order relativistic interaction vertices in our method, which is new, to the best of our knowledge. The results in the static limit was given, and the relativistic corrections to the static polarizability were derived, which are in agreement with the known published results.
The calculation of magnetic dipole polarizability was based on Appendix~\ref{appendix_note_1}. There should be no special difficulties in treating magnetic multipole interactions. The leading-order relativistic corrections to the magnetic dipole polarizability were given.

A special type of interaction due to the Coulomb-transverse photon arises naturally when the FW and PZ transformations are applied to the Dirac Hamiltonian. As we discussed in Sec.~\ref{other types of corrections}, in the case of the Coulomb-transversal photon interaction, this interaction term has no contribution from parity consideration. Nevertheless, we cannot rule out the possibility of having some non-perturbative terms that destroy the parity and thus result in a non-zero contribution. Finally, we briefly discussed the electric multipole contributions and their coupling with the magnetic dipole moment.

\section{Acknowledgement}
Authors wish to thank valuable suggestions from Z.-C. Yan at New Brunswick University and from Z.-X. Zhong at Wuhan Institute of Physics and Mathematics. This work was supported by the National Natural Science Foundation of
China (No.~11674253). X.-S. Mei was also supported by the National Natural Science Foundation of China (Nos. 11474316 and 91636216) and the Strategic Priority Research Program of the Chinese Academy of Sciences (No. XDB21020200).

\begin{appendix}
\section{A Note on the PZ Transformation}\label{appendix_note_1}
The Power-Zienau transformation applied on FW Hamiltonian is
$$
\mathcal{H}_{\rm PZ}=e^{-\imath\phi}\mathcal{H}_{\rm FW}e^{\imath\phi}+\partial _t\phi\,,
$$
where $\mathcal{H}_{\rm FW}$ is the FW Hamiltonian Eq.~\eqref{reversed_FW_hami}
\begin{equation}
\begin{aligned}
\mathcal{H}_{FW}=\frac{\left( \bm{\sigma}\cdot \bm{\Pi} \right) ^2}{2m}+eA^0-\frac{\left( \bm{\sigma}\cdot \bm{\Pi} \right) ^4}{8m^3}-\frac{\imath}{8m^2}\left[ \bm{\sigma}\cdot \bm{\Pi} ,\bm{\sigma}\cdot \bm{\mathcal{E}} \right]\,.
\notag
\end{aligned}
\end{equation}
In the above, $\bm{\Pi}=\bm{p}-e(\bm{A}_\text{at}+\bm{A}_{\text{ex}})=\bm{\pi}-e\bm{A}_{\text{ex}}$, $\bm{A}_{\text{at}}$ is the electromagnetic field from within the atom, $\bm{A}_{\text{ex}}$ is the external electromagnetic field, and $eA^0_{\text{ex}} + \partial_t \phi$ is given in Eq.~\eqref{multipole_indi}.
Here we focus on $e^{-\imath\phi}\bm{\Pi}e^{\imath\phi}$.

We first write down
\begin{equation}\label{trans1}
\begin{aligned}
e^{-\imath\phi}\bm{\Pi} e^{\imath\phi}
=e^{-\imath\phi}\left( \bm{p}-e\bm{A} \right) e^{\imath\phi}
\simeq \bm{p}
+
\left[ -\imath\phi ,\hat{p} \right]
-e\bm{A}_{\text{at}}
-e\bm{A}_{\text{ex}},
\end{aligned}
\end{equation}
in which we expand the exponential terms and reserve the first order in $\phi$.
The above expression can further be simplified by noting
$\hat{p}=-\imath \nabla$
\begin{equation}\label{trans2}
\begin{aligned}
e^{-\imath\phi}\bm{\Pi} e^{\imath\phi}
=\bm{p}-e\bm{A}_{\text{at}}+\left( \nabla \phi \right) -e\bm{A}_{\text{ex}},
\end{aligned}
\end{equation}
where $\left( \nabla \phi \right) -e\bm{A}_{\text{ex}}$ is responsible for magnetic dipole and multipole interaction.
Using the expansion of $\phi$ listed in Eq.~(\ref{A_multipole})
 yields
\begin{equation}
\begin{aligned}
&
\left[ \left( \nabla \phi \right) -e\bm{A}_{\text{ex}} \right]^m
\\
&=\nabla^m e\left[ r^iA_{\text{ex}}^{i}\left( \text{0,}t \right) +\frac{1}{\text{2!}}r^ir^jA_{\text{ex}}^{i}\left( \text{0,}t \right) _{,j}+\frac{1}{\text{3!}}r^ir^jr^kA_{\text{ex}}^{i}\left( \text{0,}t \right) _{,jk}+\frac{1}{\text{4!}}r^ir^jr^kr^lA_{\text{ex}}^{i}\left( \text{0,}t \right) _{,jkl}+\cdots \right]
\\&
-e\left[ A_{\text{ex}}^{m}\left( \text{0,}t \right) +r^iA_{\text{ex}}^{m}\left( \text{0,}t \right) _{,i}+\frac{1}{\text{2!}}r^ir^jA_{\text{ex}}^{m}\left( \text{0,}t \right) _{,ij}+\frac{1}{\text{3!}}r^ir^jr^kA_{\text{ex}}^{m}\left( \text{0,}t \right) _{,ijk}+\cdots \right]
\\&
=e\left\{ \frac{r^i}{\text{2!}} \mathcal{B}_{\text{ex}}^{mi}
+\frac{2r^i}{\text{3!}}r^j\partial _j \mathcal{B}_{\text{ex}}^{mi}
+\frac{3r^i}{\text{4!}}r^jr^k\partial _j\partial _k \mathcal{B}_{\text{ex}}^{mi}+\cdots +\frac{nr^i\left[ r^{(n-1)}:\nabla ^{(n-1)} \right]}{\left( n+1 \right) !} \mathcal{B}^{mi}_{\text{ex}}+\cdots \right\},
\end{aligned}
\end{equation}
where the superscript $m$ is a 3-dimensional spatial index. The notation $r^{(n)}:\nabla^{(n)}$ means an arrangement that $n$-number $r$ and $n$-number $\nabla$. For example, when $n=3$, this notation is given in a explicit expression $r^{(3)}:\nabla^{(3)}=r^j r^k r^l \nabla_j \nabla_k \nabla_l$. This is a general formula for magnetic multipole interaction in the Hamiltonian.
Therefore, the transformed $\bm{\Pi}$ becomes
\begin{equation}
\begin{aligned}
e^{-\imath\phi}\bm{\Pi}e^{\imath\phi}
=
\bm{\pi}+e\left[ \frac{r^i}{\text{2!}} \mathcal{B}_{\text{ex}}^{mi}+\frac{2r^i}{\text{3!}}r^j\partial _j \mathcal{B}_{\text{ex}}^{mi}+\frac{3r^i}{\text{4!}}r^jr^k\partial _j\partial _k \mathcal{B}_{\text{ex}}^{mi}+\cdots +\frac{nr^i\left( r^{n-1}:\nabla ^{n-1} \right)}{\left( n+1 \right) !} \mathcal{B}^{mi}_{\text{ex}}+\cdots \right]
\end{aligned}
\end{equation}
In our discussion only the contribution from the magnetic dipole moment is considered, because higher-order contributions are negligible in comparison with the electric interaction.

\section{A Note on Radiative Calculation}\label{appendix_note_2}
As we mentioned above, the virtual photon interaction is caused by the atomic Hamiltonian Eq.~\eqref{atomic_hami}.
If we only consider the electromagnetic field within the atom, this Hamiltonian is exactly the FW Hamiltonian. The nonrelativistic approximated interaction vertices can be constructed using this Hamiltonian, as listed in Table~\ref{table1}.
\begin{table*}
\centering
\caption{Feynman rules derived from NRQED, up to relativistic interaction. $\bm{p}$ and $\bm{p}'$ are the momenta for incoming and outgoing electrons respectively, and $\bm{q}$ and $\bm{q}'$ are the corresponding momenta for photons.}\label{table1}
\begin{tabular}{c}
\hline
\hline
Nonrelativistic vertices
\\
\hline
dipole and fermion vertex
\\
$\frac{-e}{2m} (\bm\sigma\cdot\bm{p'}\sigma^i+\sigma^i \bm\sigma\cdot\bm{p})$
\\
seagull vertex
\\
$\frac{e^2}{2m} \delta^{ij}$
\\
\hline
Relativistic vertices (first order)
\\
\hline
relativistic dipole and fermion vertex
\\
$
\frac{e}{8m^3}\left[ \sigma ^i\left( {\bm\sigma}\cdot \bm{p} \right) \left( {\bm\sigma}\cdot \bm{p} \right) ^2+\left( \bm{\sigma}\cdot \bm{p}' \right) \sigma ^i\left( \bm{\sigma}\cdot \bm{p} \right) ^2+\left( \bm{\sigma}\cdot \bm{p}' \right) ^2\sigma ^i\left( \bm{\sigma}\cdot \bm{p} \right) +\left( \bm{\sigma}\cdot \bm{p}' \right) ^2\left( \bm{\sigma}\cdot \bm{p}' \right) \sigma ^i \right]
$
\\
time-derivative vertex
\\
$
\frac{-\omega}{8m^2}\left[ \left( \bm{\sigma}\cdot \bm{p}' \right) \sigma ^i-\sigma ^i\left( \bm{\sigma}\cdot \bm{p} \right) \right]
$
\\
relativistic two-photon exchange vertex
\\
$
\frac{-e^2}{8m^3}\left[ \left\{ \sigma ^i,\sigma ^j \right\} \left( \bm{\sigma}\cdot \bm{p} \right) ^2+\left( \bm{\sigma}\cdot \bm{p}' \right) ^2\left\{ \sigma ^i,\sigma ^j \right\} +\left( \bm{\sigma}\cdot \bm{p}' \right) \left\{ \sigma ^i,\sigma ^j \right\} \left( \bm{\sigma}\cdot \bm{p} \right) \right]
$
\\
$
+\frac{-e^2}{8m^3}\left[ \sigma ^i\left( \bm{\sigma}\cdot \left( \bm{p}'+\bm{q}' \right) \right) \left( \bm{\sigma}\cdot \left( \bm{p}+\bm{q} \right) \right) \sigma ^j+\sigma ^j\left( \bm{\sigma}\cdot \left( \bm{p}'+\bm{q} \right) \right) \left( \bm{\sigma}\cdot \left( \bm{p}+\bm{q}' \right) \right) \sigma ^i \right]
$
\\
$
+\frac{-e^2}{8m^3}\left[ \sigma ^i\left( \bm{\sigma}\cdot \left( \bm{p}'+\bm{q}' \right) \right) \sigma ^j+\sigma ^j\left( \bm{\sigma}\cdot \left( \bm{p}'+\bm{q} \right) \right) \sigma ^i \right] \left( \bm{\sigma}\cdot \bm{p} \right)
$
\\
$
+\frac{-e^2}{8m^3}\left( \bm{\sigma}\cdot \bm{p}' \right) \left[ \sigma ^i\left( \bm{\sigma}\cdot \left( \bm{p}+\bm{q} \right) \right) \sigma ^j+\sigma ^j\left( \bm{\sigma}\cdot \left( \bm{p}+\bm{q}' \right) \right) \sigma ^i \right]
$
\\
\hline
\hline
\end{tabular}
\end{table*}
Recall the $\Sigma$ operators in Eq.~\eqref{virtual-photon-loop}, which contain the interaction vertices $p^i/m$.
In our calculation, this single-photon nonrelativistic interaction vertex is called dipole and fermion vertex in Table~\ref{table1}. The first order of relativistic vertices are also listed in the Table, though we did not use them in this work.

Next we demonstrate how to eliminate the divergence in our calculation. As mentioned above, the radiative correction to the energy shift is divided into high- and low-energy parts. An artificial factor $K$ is introduced to treat the divergence due to the virtual-photon loop. The $K$-dependent terms in low- and high-energy parts should cancel out with each other.
We first consider the low-energy part. As introduced in Eq.~\eqref{counter-operator}, the irreducible interaction operators $\Sigma$ can be divided into $\Sigma^{\ln K}$ and $\tilde{\Sigma}$, where $\Sigma^{\ln K}$ represents the divergent part.
Then the low-energy divergent part can be written in the form
\begin{equation}\label{low-energy-lnk-b}
\begin{aligned}
\left< \delta H_{\rm E,QED}^{{\rm L},\ln K} \right>
&=\left< \Sigma _{0}^{\ln K}\frac{Q}{E_0-H_0}\delta H_{\rm E,NR}+\delta H_{\rm E,NR}\frac{Q}{E_0-H_0}\Sigma _{0}^{\ln K} \right>
\\&
+e^2\mathcal{E}^i_{\rm ex}\mathcal{E}^{j\dag}_{\rm ex}\left< \left[ r^iG_0\left( -\omega \right) \Sigma _{0}^{\ln K}G_0\left( -\omega \right) r^j+r^jG_0\left( \omega \right) \Sigma _{0}^{\ln K}G_0\left( \omega \right) r^i \right] \right>
\\&
+\mathcal{E}^i_{\rm ex}\mathcal{E}^{j\dag}_{\rm ex}\left< \left[ \Sigma _{1}^{i,\ln K}G_0\left( -\omega \right) \left( -er^j \right) +\Sigma _{1}^{j,\ln K}G_0\left( \omega \right) \left( -er^i \right) \right] \right>
\\&
+\mathcal{E}^i_{\rm ex}\mathcal{E}^{j\dag}_{\rm ex}\left< \left[ \left( -er^i \right) G_0\left( -\omega \right) \Sigma _{1}^{j,\ln K}+\left( -er^j \right) G_0\left( \omega \right) \Sigma _{1}^{i,\ln\text{K}} \right] \right>
\\&
+\left< \delta H_{\rm E,NR} \right> \left( \partial _{E_0}\Sigma _{0}^{\ln K} \right) +\left< \Sigma _{0}^{\ln K} \right> \left( \partial _{E_0}\delta H_{\rm E,NR} \right)\,.
\end{aligned}
\end{equation}
Using the following identity
$$
\frac{{p}^k}{m}\left( E_0-H_0 \right) \frac{{p}^k}{m}=\frac{1}{2}\left\{ \frac{{p}^k}{m}\left[ \left( E_0-H_0 \right) ,\frac{{p}^k}{m} \right] +\left[ \frac{{p}^k}{m},\left( E_0-H_0 \right) \right] \frac{{p}^k}{m}+\frac{{p}^k}{m}\frac{{p}^k}{m} \right\}\,,
$$
the low-energy divergent part becomes
\begin{equation}
\begin{aligned}
\left< \delta H_{\rm E,QED}^{{\rm L},\ln K} \right>
&=
\frac{-\alpha}{3\pi m^2}\ln 2K\left< \left[ \nabla ^2V\frac{Q}{\left( E_0-H_0 \right)}\delta H_{\rm E,NR}+\delta H_{\rm E,NR}\frac{Q}{\left( E_0-H_0 \right)}\nabla ^2V \right] \right>
\\&
+\frac{2\alpha}{3\pi m^2}\ln 2K\left( \mathcal{E}_{\text{ex}}^{i}\mathcal{E}_{\text{ex}}^{j\dag} \right) \left< r^iG_0\left( -\omega \right) \left( \nabla ^2V-\left< \nabla ^2V \right> \right) G_0\left( -\omega \right) r^j \right>
\\&
+\frac{2\alpha}{3\pi m^2}\ln 2K\left( \mathcal{E}_{\text{ex}}^{i}\mathcal{E}_{\text{ex}}^{j\dag} \right) \left< r^jG_0\left( \omega \right) \left( \nabla ^2V-\left< \nabla ^2V \right> \right) G_0\left( \omega \right) r^i \right>
\end{aligned}
\end{equation}

The high-energy part of radiative correction is given in Eq.~\eqref{high-energy}.
By taking the term containing $\ln K$ from the QED Hamiltonian in Eq.~\eqref{QED-hamiltonian}, we have
\begin{equation}
\begin{aligned}
H_{\rm QED}^{\ln K}=\frac{\alpha}{3\pi m^2}\ln \left(\frac{m}{2K}\right) \nabla ^2V\left( r \right)\,.
\end{aligned}
\end{equation}
Substituting this into Eq.~\eqref{high-energy} yields the logarithmic term of the high-energy part
\begin{equation}
\begin{aligned}
\left< \delta H_{\rm E,QED}^{{\rm H},\ln K} \right>&=\left< H_{\rm QED}^{\ln K}\frac{Q}{E_0-H_0}\delta H_{\rm E,NR}+\delta H_{\rm E,NR}\frac{Q}{E_0-H_0}H_{\rm QED}^{\ln K} \right>
\\&
+e^2\mathcal{E}_{\text{ex}}^{i}\mathcal{E}_{\text{ex}}^{j\dag}\left< \left[ r^iG_{\rm QED}^{\left( 1 \right) \ln K}\left( -\omega \right) r^j+r^jG_{\rm QED}^{\left( 1 \right) \ln K}\left( \omega \right) r^i \right] \right>,
\end{aligned}
\end{equation}
where $
G_{\rm QED}^{\left( 1 \right) \ln K}\left( \pm \omega \right) =-G_0\left( \pm \omega \right) \left( H_{\rm QED}^{\ln K} - \langle H_{\rm QED}^{\ln K} \rangle \right) G_0\left( \pm \omega \right)
$. Because of $H_{\rm QED}^{\ln K}$ containing $\nabla^2 V(r)$, one can see that
the divergent parts in both
low- and high-energy expressions cancel out completely.

%
%
%
\end{appendix}
\bibliography{Rel-polarizability}
\end{document}